\documentclass[%
 onecolumn,
showpacs,preprintnumbers,
 amsmath,amssymb,
 aps,
 pre,
]{revtex4-1}

\usepackage[pdftex]{graphicx}
\usepackage{dcolumn}
\usepackage{bm}
\usepackage{hyperref}

\usepackage{subfigure}
\usepackage{amsmath, amssymb}
\usepackage{enumitem}
\usepackage{color, colortbl}
\usepackage{afterpage}
\usepackage{url}
\usepackage{multirow}
\usepackage{rotating}
\usepackage{scalefnt}
\usepackage{hyperref}
\usepackage{ulem}
\usepackage{braket}
\usepackage{longtable}
\usepackage{xr}
\usepackage{soul}

\begin{document}

\title{Planning for sustainable Open Streets in pandemic cities}

\author{Daniel Rhoads}
\affiliation{Internet Interdisciplinary Institute (IN3), Universitat Oberta de Catalunya, Barcelona, Spain}

\author{Albert Sol\'e-Ribalta}
\affiliation{Internet Interdisciplinary Institute (IN3), Universitat Oberta de Catalunya, Barcelona, Catalonia, Spain}
\affiliation{URPP Social Networks, Universit\"at Z\"urich, Z\"urich, Switzerland}

\author{Marta C. Gonz\'alez}
\affiliation{Department of City and Regional Planning, University of California, Berkeley, CA 94720, USA}
\affiliation{Energy Technologies Area, Lawrence Berkeley National Laboratory, Berkeley, CA 94720, USA}
\affiliation{Department of Civil and Environmental Engineering, University of California, Berkeley, CA 94720, USA}

\author{Javier Borge-Holthoefer}
\affiliation{Internet Interdisciplinary Institute (IN3), Universitat Oberta de Catalunya, Barcelona, Catalonia, Spain}

\date{\today}

\begin{abstract}
In the wake of the pandemic, the inadequacy of urban sidewalks to comply with social distancing \cite{gordon2020,fleischer2020,sadikkhan2020} remains untackled in academy. Beyond isolated efforts (from sidewalk widenings to car-free Open Streets) \cite{india2020, guerrero2020}, there is a need for a large-scale and quantitative strategy for cities to handle the challenges that COVID-19 poses in the use of public space. The main obstacle is a generalized lack of publicly available data on sidewalk infrastructure worldwide, and thus city governments have not yet benefited from a complex systems approach of treating urban sidewalks as networks. Here, we leverage sidewalk geometries from ten cities in three continents, to first analyze sidewalk and roadbed geometries, and find that cities most often present an {\it arrogant} distribution of public space \cite{arrogance2018,creutzig2020fair}: imbalanced and unfair with respect to pedestrians. Then, we connect these geometries to build a sidewalk network --adjacent, but not assimilable to road networks, so fertile in urban science \cite{Batty2012,barthelemy2018morphogenesis}. In a no-intervention scenario, we apply percolation theory to examine whether the sidewalk infrastructure in cities can withstand the tight pandemic social distancing imposed on our streets. The resulting collapse of sidewalk networks, often at widths below three meters, calls for a cautious strategy, taking into account the interdependencies between a city's sidewalk and road networks \cite{radicchi2015percolation}, as any improvement for pedestrians comes at a cost for motor transport. With notable success, we propose a shared-effort heuristic that delays the sidewalk connectivity breakdown, while preserving the road network's functionality.
\end{abstract}

\maketitle

\section{Introduction}

As the search for effective treatment regimes or a successful vaccine struggles ahead \cite{Graham2020,Thorp2020,le2020covid,callaway2020coronavirus}, the coronavirus disease 2019 (COVID-19) pandemic continues to spread worldwide. In the absence of therapeutic countermeasures, one of the most successful recommendations (interventions) has been to advise physical distancing (often referred to as ``social distancing'') in order to minimize person-to-person transmission \cite{Mossong_2008,hsiang2020}. On the other hand, stringent social distancing (e.g. shelter-in-place orders, or school and business closures) implies economic and societal stresses which become unsustainable for the good functioning of the system. In the context of the urban built environment, these antagonistic needs pose a great challenge. City-dwellers are having to learn on-the-fly how to move around in the public space of the city, while at the same time keeping a distance of at least 1.5 meters from their fellow citizens \cite{chu2020physical,WHO,streets2020Nacto}. Concomitant factors --a precipitous drop in public transit ridership \cite{troko2011public,goldbaum2020,batty2020coronavirus}, the possible relationship between pollution and mortality \cite{wu2020exposure,stier2020covid}-- seem to leave walking and cycling as the only transport options suitable for maintaining satisfactory levels of health and well-being \cite{deVos2020} in shared public space.

While the term public space may conjure up images of parks and greenways, it is easy to overlook perhaps the most important public space of all: the sidewalk. In the wake of the pandemic, popular media have given much  attention to the inadequacy of social distancing on our cities' sidewalks \cite{gordon2020, fleischer2020, sadikkhan2020}. Many cities have followed suit, implementing {\it ad hoc} interventions (from temporary sidewalk widenings to the establishment of car-free ``Open Streets") \cite{india2020, guerrero2020} to give pedestrians more space, measures which serve both to decrease the probability of contagion \cite{hsiang2020} and to provide people with a sense of relief in terms of perceived risk \cite{batty2020coronavirus,seale2020covid,Mkel2020}. So far, these interventions have been local and manual, and have not directly benefited from a complex systems approach of treating urban sidewalks as a network, in part due to a due to a generalized lack of publicly available data on sidewalk infrastructure worldwide.
	
In this work, we bridge this initial data gap by collecting comprehensive datasets of sidewalk and road infrastructure from ten world cities across three continents. First, we quantify the share of public space allotted to pedestrians versus cars, taking the measure of {\it arrogance of space} as defined by Colville-Andersen \cite{arrogance2018} to a city-wide level and showing for the first time on such a large scale how planning choices can leave pedestrians with less room to walk. 
In addition to quantifying the need for space for pedestrians, it is also important to measure the overall connectivity of the sidewalk infrastructure. Thus, building from the same sidewalk geometry data, we develop a method to automatically construct city-scale sidewalk networks, which we apply to our cities of study.
	
Next, we assess the potential decline in connectivity suffered by our empirical sidewalk networks as a consequence of current social distancing recommendations, leveraging tools from the field of complex networks. Up to now, nearly no quantitative analysis exists regarding the robustness of sidewalk networks, with or without social distancing considerations. Anchoring our analysis to the World Health Organization (WHO) \cite{WHO} and National Association of City Transportation Officials (NACTO) recommendations \cite{nacto2013urban}, we provide a baseline from which to adjust urban infrastructure, considering the delicate trade-offs between the sidewalk and road networks. Our proposal can either be applied in the context of the pandemic --as an extraordinary and temporary intervention--, or as a long term strategy to rebalance the public space. 


\begin{figure}[h]
\centering
\includegraphics[width=\textwidth]{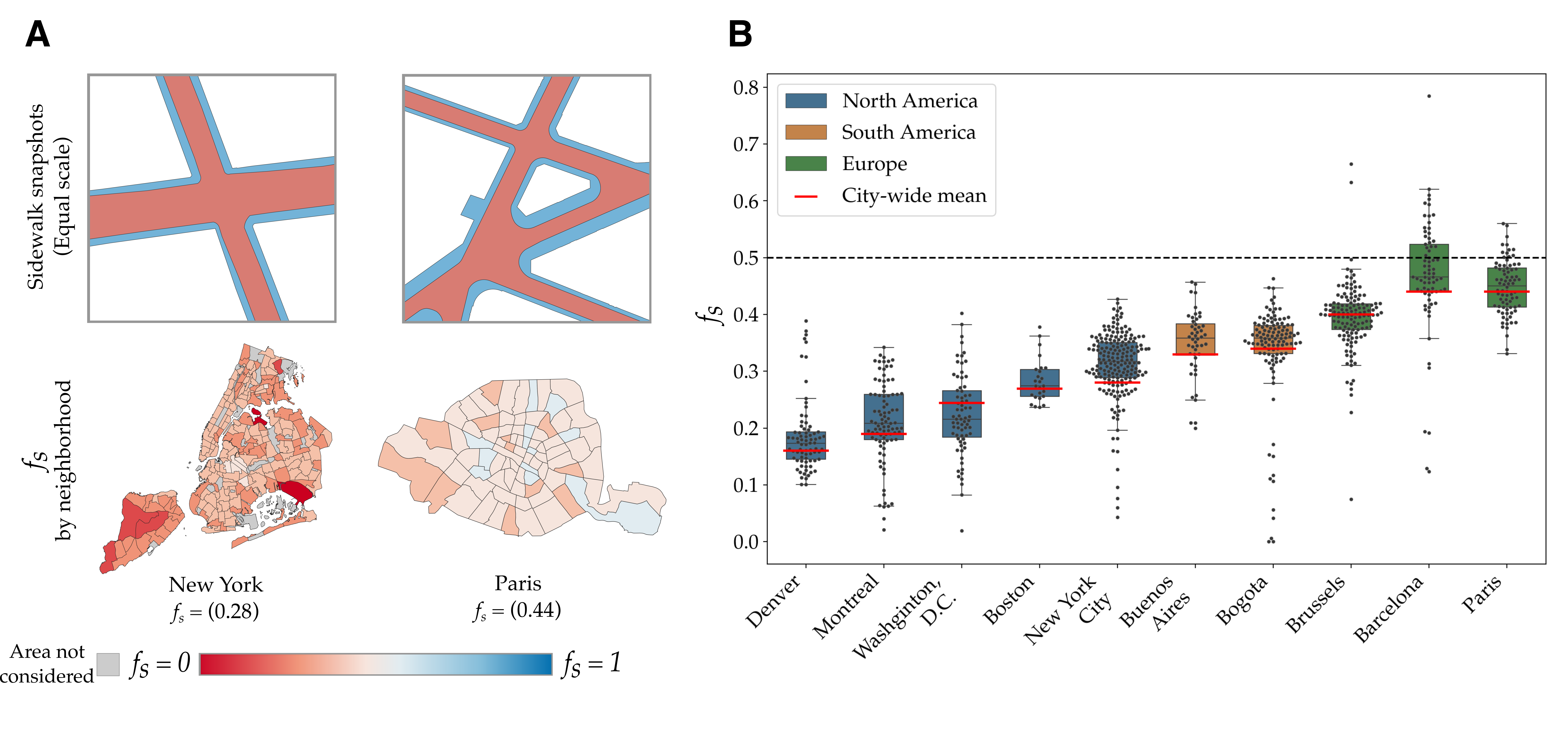}
\caption{{\bf Maps and statistics on Arrogance of Space.} The distribution of street space between cars and pedestrians varies widely both within and among cities, but is everywhere biased towards cars. {\bf A} On top, close-up snapshots representing the area for motor vehicles (red) and sidewalks (blue). White area corresponds to buildings. Clear to the naked eye, the majority of street space is devoted to cars. A more general view of the cities under scrutiny (bottom) confirms indeed that roads take up most of the space. Only Paris presents a few districts where the sidewalk share of street space $f_s$ is above 50\%. {\bf B} Box plots displaying the distribution of space all the cities for which the sidewalk geometries were gathered. The x-axis is sorted from lowest to highest city-wide $f_s$. Interestingly, none of the cities achieve a median (nor average) $f_s$ of 50\% or higher. In general, cities present a remarkable internal heterogeneity, which implies that some district suffer from a double inequity (with respect to cars, and also with respect to other areas of the city). It is clear that European cities rank best in the plot, possibly because they typically have larger ``historical centers'', where street space tends to be shared.}
\label{fig:arrogance}
\end{figure}

\section{From pedestrian space to sidewalk networks}


As a first step to quantify urban pedestrian space, we have collected geodatasets from ten cities distributed across three continents, each comprised of a set of \textit{road} and \textit{sidewalk} information representing the geographic extent of those two features (see Supplementary Materials for more information). From this data, we can map out the space allocated to cars and pedestrians respectively, both for the city as a whole and within specific districts (see Figure~\ref{fig:arrogance}). Even at a small scale (panel A, top), it is already apparent that vehicle space takes most of street space in the scenes. Scaling up this analysis to the city-wide level (panel A, bottom), we see that even in a best-case scenario (Paris, France), sidewalks occupy just under half of the available space ($f_{s} = 0.44$); at the other extreme, New York devotes as little as 28\% of its public space to sidewalks. Moreover, these fractions are not homogeneously distributed across the urban landscape. Indeed, the arrogance of space \cite{arrogance2018} is, in general, aggravated as we move out from the city centre, as can be seen in the case of New York where darker reds are concentrated in the peripheral areas of the outer boroughs. However, exceptions do arise: in Paris, for example, districts with a higher share of sidewalk space are more evenly distributed throughout the city, possibly indicating enclaves of older, more walkable neighborhoods amid modernized, car-friendly areas.

Moving from a city-wide to a global scale, the box plots in Figure~\ref{fig:arrogance} (panel B) summarize the district-level street space distributions across all 10 cities of study. In the lower end of the distributions, a few cities present an alarming lack of sidewalk infrastructure, with some districts allocating just 10\% of their street space to sidewalks. With the exceptions of Boston and Paris, which both present relatively compact distributions of sidewalk share, most of the cities exhibit large heterogeneities in their distributions of the space devoted to pedestrians. Note that the box plots in panel B of Figure~\ref{fig:arrogance} have been organized in ascending order, taking the average fraction of sidewalk in each city as a criterium. Remarkably, this results in a natural ordering of the cities by continent.

This initial analysis provides a glimpse of how street space is currently distributed in cities around the world, and allows us to hypothesize that social distancing measures could pose a variety of challenges, both within cities and among countries and continents. Validating this requires us to look beyond share of space dedicated to pedestrians at a specific location, to the service that sidewalks provide as urban transportation infrastructure. To this end, it is necessary to translate sidewalk geometries into a connected structure: a sidewalk network. 

\begin{figure}[h!]
\centering
	\includegraphics[width=.80\textwidth]{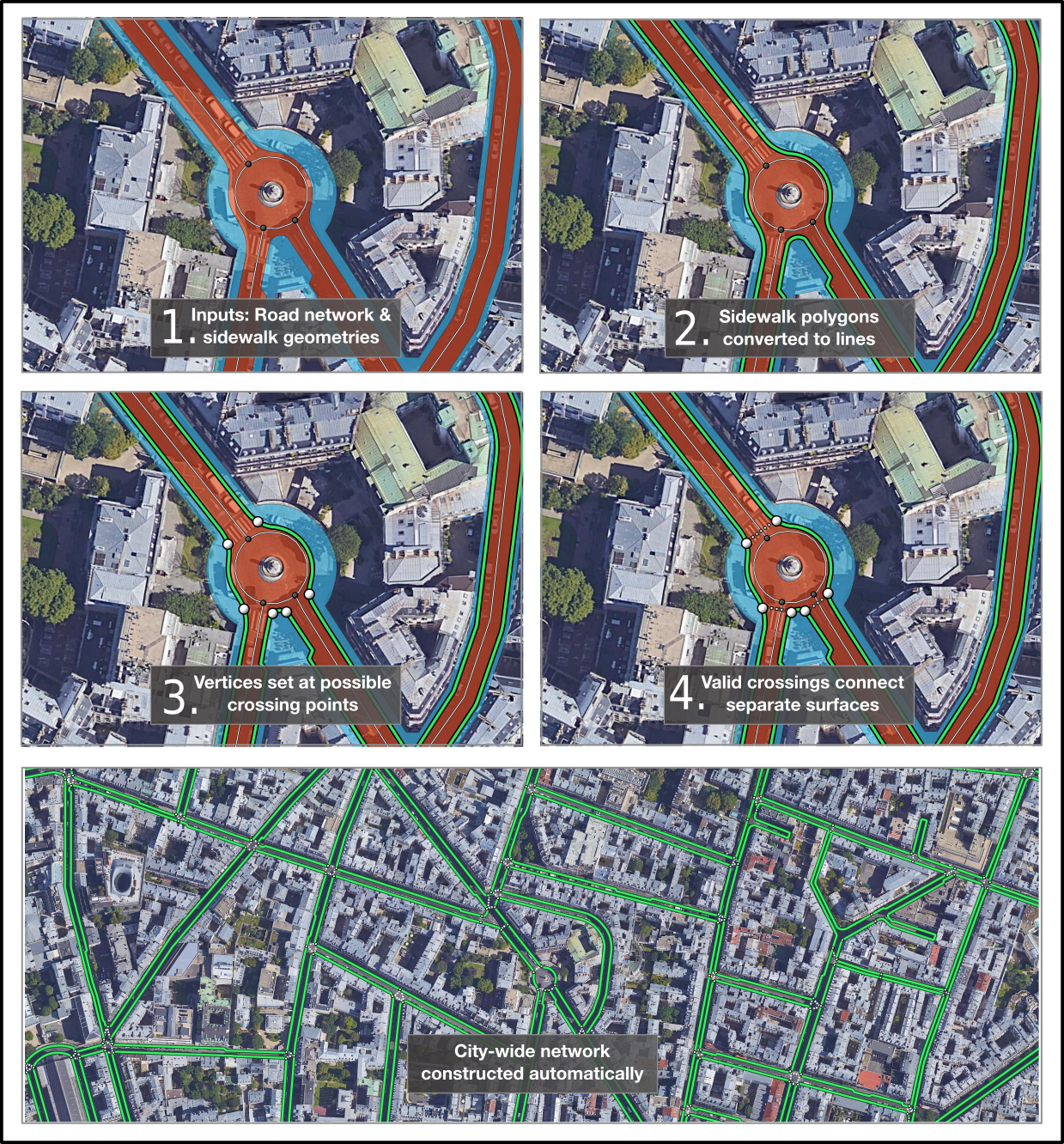}
\caption{{\bf Network construction process.} Building from heterogeneous municipal open data sets, an algorithm was developed to construct sidewalk networks automatically for each of the cities of study using a few simple rules, as described in the figure.}
\label{fig:construction}
\end{figure}

Building mainly from planimetric data, we follow a minimalistic (yet accurate) approach to sidewalk network construction, which is illustrated in Figure~\ref{fig:construction}. First, a node is placed whenever pedestrians have the choice either to change direction, or to change the surface on which they are walking. As such, edges are laid to link two nodes on the same surface (e.g. two corners of a block), or else to link the closest node across the street (e.g. a crosswalk). The resulting structure is an undirected, spatially-embedded, weighted network, where edges can be assigned a variety of relevant attributes (length, width, slope, etc.). A detailed description of the network construction process can be found in the Supplementary Materials.

\begin{figure}[h!]
\centering
\includegraphics[width=\textwidth]{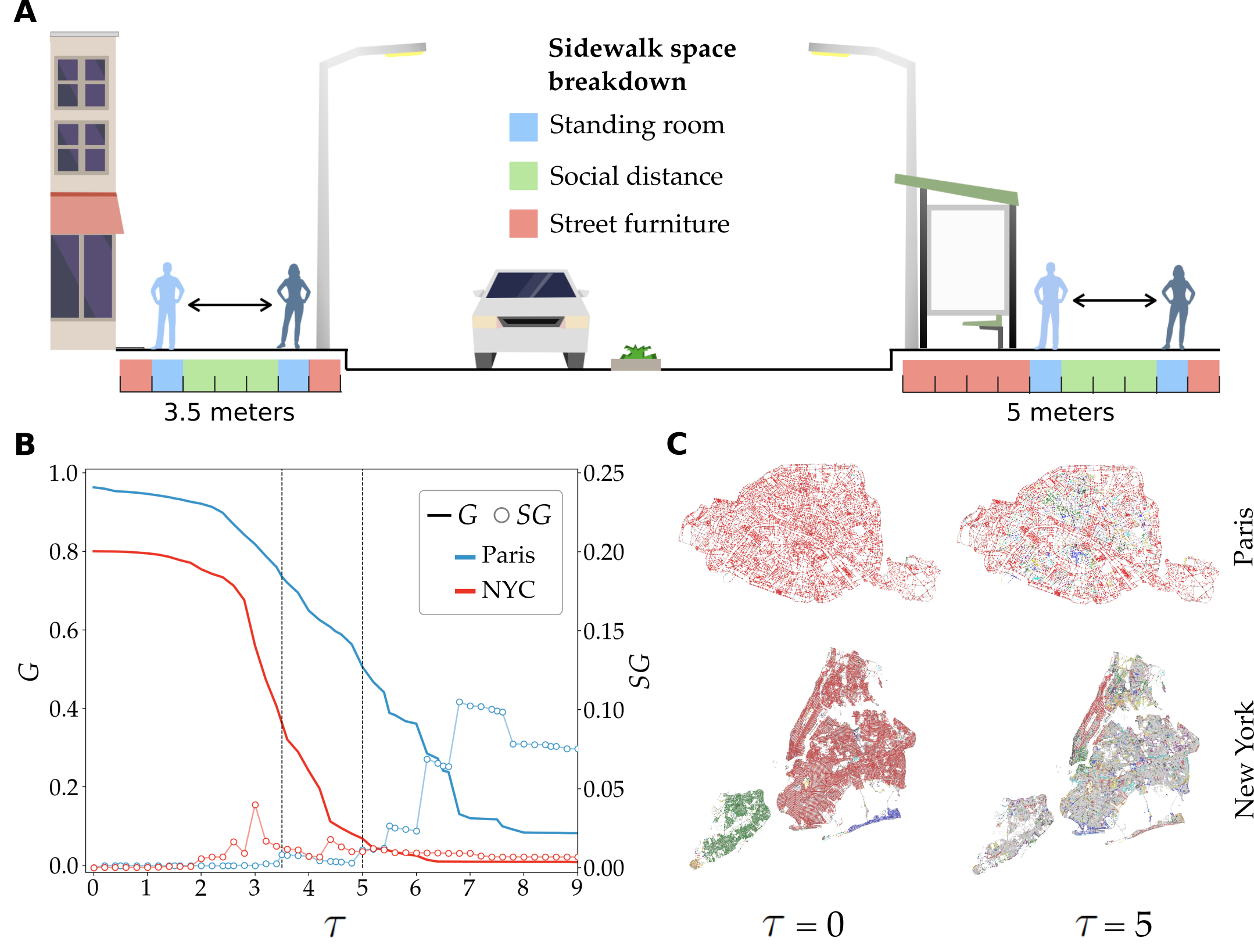}
\caption{{\bf The implications of social distancing guidelines on sidewalk connectivity.} Panel A on top illustrates two social distancing scenarios. In one (3.5 meters), two people may walk comfortably keeping 1.5 meters of physical distance, with only 1 meter left for street furniture; in the other, street furniture occupies 2.5 meters. Panel B illustrates the change in the size of the giant component $G$ of the sidewalk network of New York City (red) and Paris (blue) as sidewalks are iteratively removed, from narrowest to widest. The evolution of the second giant component $SG$ is represented as well. While Figure~\ref{fig:arrogance} shows that the two cities dedicate different amounts of space to pedestrians, neither network can withstand the stringent requirements of 5 meters. Finally, panel C shows the breakdown of both cities from their original state ($\tau = 0$), to their state when all sidewalks below 5 meters in width have been removed. Sidewalk nodes are colored according to the connected component to which they belong.}
\label{fig:paris}
\end{figure}

\section{Social distancing on sidewalk networks}
With our newly-constructed sidewalk networks at hand, we can proceed to engage the problem of social distancing on sidewalks in a systematic way, as has been done with respect to other contexts such as public transportation \cite{troko2011public,deVos2020}. In this section, we first test the networks' present capability to provide for recommended safe distancing. We then go on to propose an automated heuristic to improve the current situation, emphasizing sidewalk connectivity, as well as trade-offs with the urban road network. 

\subsection{Status quo}

The natural approach to assessing the feasibility of sidewalk social distancing from a network perspective is through targeted edge percolation \cite{albert2000error,cohen2001breakdown,abbar2018structural}. Applied to our problem, targeted edge percolation translates to blocking sidewalks that cannot guarantee a prescribed social distance by removing all links $l$ whose width falls below a predefined threshold of $\tau$ meters, i.e. $l \le \tau$. In other words, the threshold $\tau$ is applied in ascending order of width.

As $\tau$ increases, we monitor the size of the network's giant connected component $C_1$ as a fraction of the size of the network $N$, defined as $C_1/N = G$, which informs us to what extent the network is still functional, i.e. navigable. As a reference, we will focus on 3.5- and 5-meter width sidewalks, which can be regarded as the minimal width necessary for two people to comfortably walk side-by-side on a sidewalk while maintaining social distancing recommendations (1.5 meters) \cite{WHO}, considering sidewalks with and without urban furniture (benches, trees, street lighting, etc.) \cite{nacto2013urban}, see Figure~\ref{fig:paris}A. 

Correspondingly, we represent two dashed vertical lines at $\tau = 3.5$ and $\tau = 5$ meters in Figure~\ref{fig:paris}B. Following the decline of $G$, we see how the sidewalk networks in Paris (blue) and New York (red) break down, as we iteratively block sidewalks of width below a threshold of $\tau$ meters. Notably, New York already presents a fragmented scenario at $\tau = 0$, where $G = 0.8$, due to its dispersal across water bodies. This implies that New York's urban sidewalk infrastructure presents discontinuities: parts of the city are unreachable on foot (this feature is not unique to New York, see Figures S1-S8 and accompanying text in the Supplementary Materials).


New York's sidewalk network rapidly reaches the critical percolation point (as indicated by the peak of the second giant component $SG = C_2/N$, circles in the plot) \cite{janson1993birth,molloy1998size} after a limiting width of $\tau \approx 3$ meters, and heavily deteriorates thereafter. At a threshold of $\tau= 5$ meters the network has totally collapsed. Paris shows a much slower rate of decline: its giant component still holds 50\% of the sidewalk network together at $\tau = 5$. 
The maps in Figure~\ref{fig:paris}C present two snapshots of the percolation process, at $\tau = 0$ (no social distance restrictions) and $\tau = 5$ (with social distance restrictions). Visible to the naked eye, Paris still exhibits a large connected component (red), but some isolated areas emerge even at the center of the city. On the other hand, only Manhattan retains a relatively functional walking infrastructure at $\tau = 5$, while most of the city's sidewalks are impracticable at that threshold.

\subsection{``Open Streets'' on interdependent networks}

Treating the problem of sidewalk social distancing from a network perspective allows us to propose improvements to the {\it status quo} that take into account the system-wide effects of any local intervention.  All of the street pacification \cite{brindle1991traffic,habitat2016united} solutions currently being implemented by cities on an {\it ad hoc} basis, from sidewalk widenings to traffic-free ``Open Streets", involve redistributing limited street space from cars to pedestrians. This creates a sensitive situation: the sidewalk and road networks of a city are coupled, and their interdependency shapes the conditions for their simultaneous operation \cite{Karndacharuk_2014,radicchi2015percolation}. Thus, while closing a street to traffic may improve the robustness of the sidewalk network, it necessarily comes at a cost for the road network. This interplay --any change in one network affects the other-- demands a balanced approach: one that guarantees the sustainability of both structures. In this context, network analysis provides a key tool to design strategies that may help determine how many (and which) streets can be pacified, while avoiding the collapse of the road network's functionality.

Among many possibilities, we propose an ``Open Streets'' greedy heuristic in which, for each $\tau$ in ascending order, we select the $\tau$-width sidewalk and the corresponding adjacent road segment. We then consider blocking that sidewalk, or blocking the neighboring roadbed. To do so, we evaluate which structure suffers least, in terms of connectivity loss (giant component relative size), when the corresponding edge is removed. If the greatest loss is for the sidewalk network, we pacify that street, blocking car access (road edge removal); if the greatest loss is for the road network, the corresponding $\tau$-wide sidewalk segment is removed. In the frequent case of a tie (same loss for both interdependent structures), then the sidewalk edge is removed with probability $p \sim B_{ij}$, where $B_{ij}$ is the time-weighted edge betweenness of the adjacent road segment $(i,j)$ (see Supplementary Materials); conversely, the road segment $(i,j)$ is removed with probability $1-p$. The {\it rationale} behind this strategy is that high-betweenness road edges are key to preventing congestion \cite{Guimera2002,sole2018decongestion,sole2019effect}, and to maintaining the travel time as low as possible. By deciding ties according to this criterium, we set a bias towards preserving high-betweenness road segments alongside the more general goal of sustaining connectivity.

\begin{figure}[h!]
\centering
\includegraphics[width=0.95\textwidth]{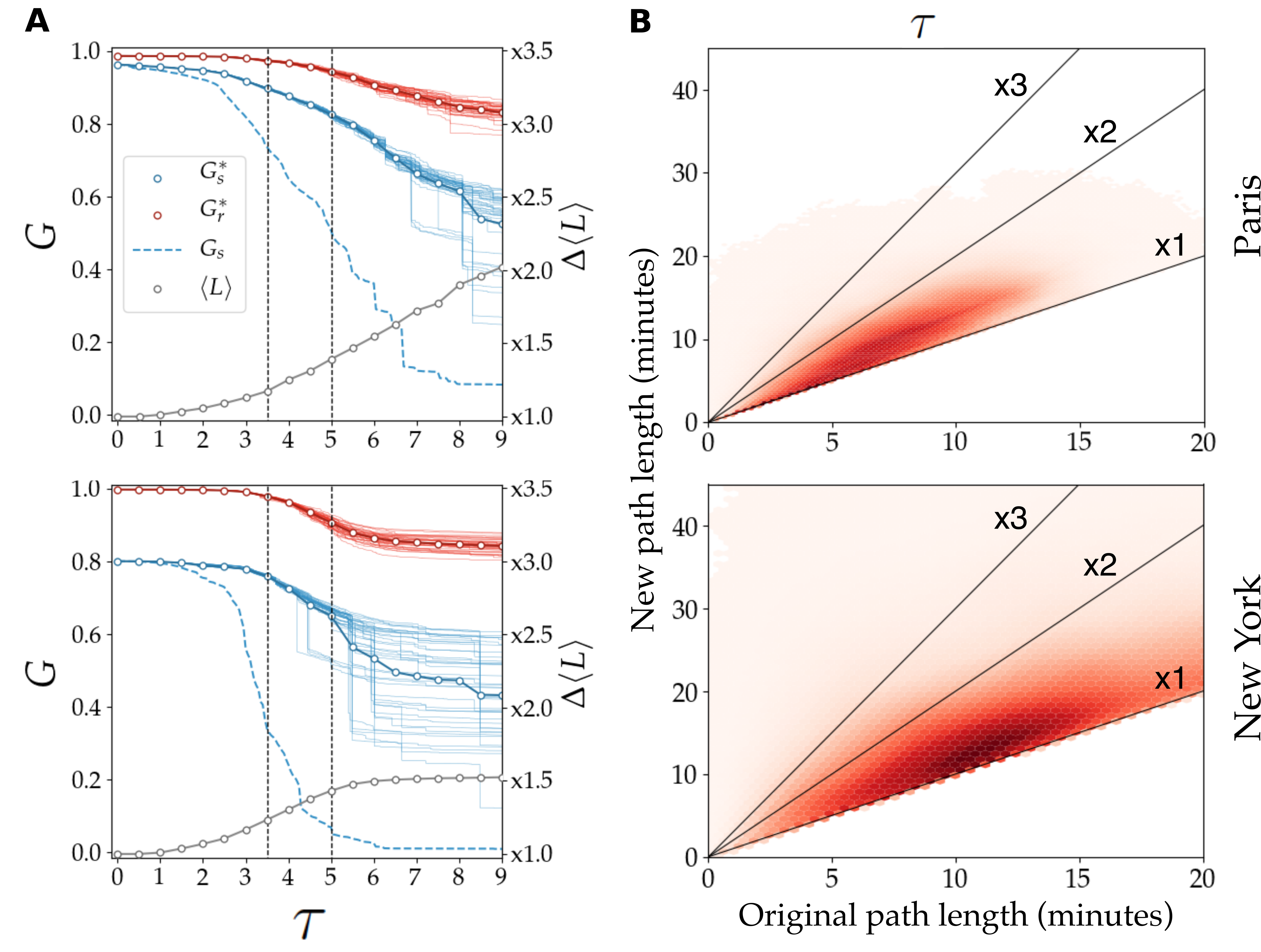}
\caption{{\bf Results of the ``Open Streets'' shared-effort heuristic.} As can be seen in the left panels, applying the process to New York City and Paris leads to significant gains for the sidewalk network with respect to the case of no intervention (blue dashed line). Additionally, the road network begins to lose some connectivity as higher-width sidewalks are pedestrianized, but these losses are relatively small. 
The right panels show the distribution of the increase in average path lengths (in minutes) for drivers when the process has been run up to sidewalks of 5 meters in width. While removed streets from the road network clearly implies longer travel times for cars, these plots show quite forcefully that the costs are relatively low and concentrated around short-to-medium paths.}
\label{fig:intervention}
\end{figure}


Figure~\ref{fig:intervention} describes this process for Paris and New York (left column). The dashed line reproduces the evolution of $G_{s}$ if no intervention takes place ($G$ in Fig.~\ref{fig:paris}), as a baseline reference. The blue and red solid lines represent, respectively, the evolution of the sidewalk and road networks' giant component under our proposed heuristics, $G*_{s}$ and $G*_{r}$ respectively, in which the effort to sustain both networks' functionality is shared. Notably, the collapse of the sidewalk network is delayed significantly, with 80\% (Paris) and 70\% (New York) of the sidewalk network still connected at a thresholding width of 5 meters. Furthermore, such delays are compatible with little harm to the integrity of the road network, whose giant component $G*_{r}$ still connects over 90\% of nodes in its structure, for both cities. Note that this strategy does not imply that {\it any} segment below 5 meters width has been pacified, but just enough that both structures could endure. The process, however, has unavoidably damaged to some extent the road network: besides a small loss in connectivity, the gray line in the plot monitors how much travel time has, on average, increased because of street pacification ($\Delta \langle L \rangle$). It turns out that, both in Paris and New York city, travel times have increased by no more than a factor of 1.5, if we set a limit at $\tau = 5$ meters. Fine-graining this aggregate view (to the level of individual path duration distribution, see the hex-binned panels to the right), we see that most of these trips span, in the non-intervened network, 10 to 15 minutes or less, which implies that, in most cases, the final travel time comes at an acceptable increased cost. Figures~S1-S8 in the Supplementary Materials reproduce the same results for the other eight cities studied here. While the shared-effort strategy renders an improvement for all cities, we observe that the heuristic in North American cities (especially in Boston, Denver, and Montreal) struggles to maintain a pedestrian giant connected component at 50\% beyond a $\tau = 3$ meters sidewalk width, although the increased cost in car travel times affects mostly short trips. This may be explained by the specific correlations that exist between the distribution of sidewalk widths and the betweenness centrality of the corresponding street segments, see Figure~S9.

\begin{figure}[h]
\centering
\includegraphics[width=0.95\textwidth]{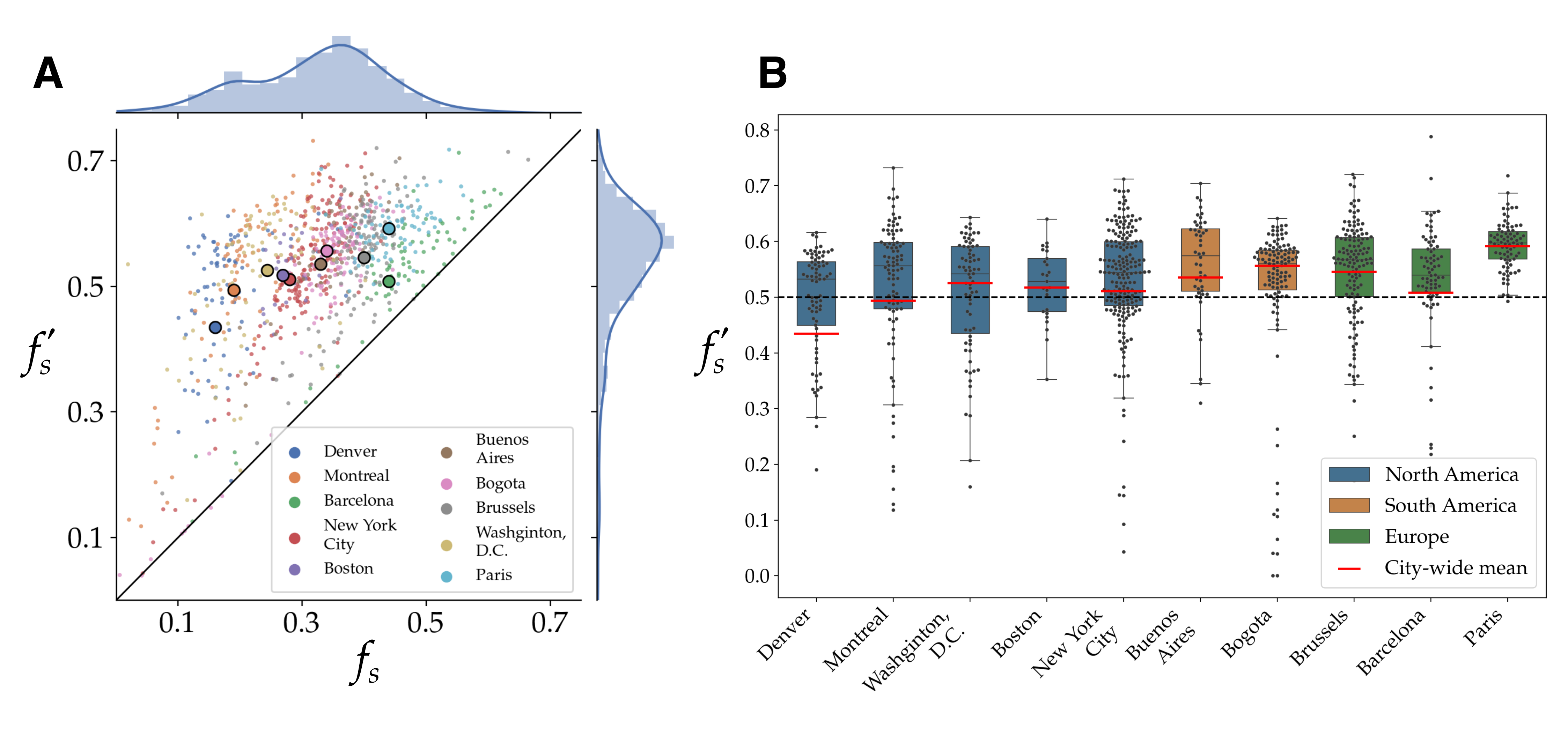}
\caption{{\bf Distribution of public space, post-intervention fixing $\tau =5$.} The scatterplot to the left compares the amount of sidewalk share before and after the proposed intervention. Clearly, pedestrianizing streets to promote social distancing also has an effect on the overall distribution of space between pedestrians and cars, as illustrated here (note the diagonal with slope 1 as a guideline to the eye). The box-plots in the right panels reproduce those of Fig.~\ref{fig:arrogance} in the same order, now considering the gains in sidewalk share.}
\label{fig:dearrogance}
\end{figure}

The proposed heuristic leads, as we have seen, to notable success both at preventing an early collapse of both street (sidewalk and road) networks, and at maintaining travel times at a reasonable level. Also, as a necessary side-effect, street pacification also alters street space share between sidewalks and roads. Figure~\ref{fig:dearrogance} repeats the results shown in Fig.~\ref{fig:arrogance} for the 10 cities under study. The scatter plot in panel A plots the post-intervention sidewalk space share $f'_s$ against the original $f_s$ for each district (smaller dots), as well as global city averages (larger dots). The marginal distributions pre- (top) and post-intervention (right) are also shown, evidencing a notable shift towards balanced road-sidewalk space share. Panel B shows the new distribution of space share in each city, preserving the order as they were laid out in Fig.~\ref{fig:arrogance}. Clearly, the post-intervention scenario is rather compact, with most cities --and districts within-- presenting a balanced $f'_s$ at $\sim$50\%.

\section{Discussion}

The disruptions to daily life provoked by the pandemic reach far beyond purely medical ones. As a consequence, research related to the virus' consequences has spilled outside epidemiology and public health, and into economics, and other social and behavioral sciences. Nevertheless, despite the significant public attention given to the issue, it is difficult --if not impossible-- to find a large-scale, quantitative assessment to the way in which cities may handle the challenges that COVID-19 poses in the use of public space. Such a gap is not surprising: a lack of wide-spread data and standardized methods has prevented the development of a literature on the subject comparable to that of their counterpart road networks, making sidewalk networks perhaps the most neglected piece of our urban transportation infrastructure --long before the pandemic emergence. It is no wonder, then, that this work is the first to address, in a systematic and city-wide manner, the inadequacy of our cities' sidewalk infrastructure for the requirements of social distancing in the context of the ongoing pandemic.

This inadequacy is already apparent from the large-scale evidence of the imbalanced distribution of public space. The informal impression that most street space is devoted to cars is accurately confirmed for a diverse selection of cities which, we suspect, are good representatives of the situation elsewhere --at least with regard to large urban areas. Connecting these sidewalk geometries from two simple rules (change in direction or surface), we construct city-wide, richly-attributed sidewalk networks. These are represented as undirected, spatially-embedded, weighted structures, which can now be analysed within the framework of urban complex networks. A targeted attack width-based percolation process carried on the cities of study reveals that current sidewalk infrastructure becomes, if no intervention is in place, severely fragmented when stringent social distancing recommendations are taken into account, with many parts of the city mutually inaccessible on foot.

And yet, city governments have and exercise the power to pacify streets, even at the cost of increasing inefficiencies for drivers. Without the underlying structure of the sidewalk network, however, it is not possible to assess the consequences of manually selected interventions at the neighborhood or city-wide levels. However, we highlight that road networks must be treated with care in order not to collapse vehicle mobility in cities. Street pacification thus demands that careful attention be given to the interdependencies between a city's sidewalk and road networks, making a systemic, data-driven strategy necessary. To this end, we propose an heuristic that emphasizes the connectivity of both the sidewalk and road networks, as well as the efficiency of the latter, with considerable gains in each of the cities studied. Also, an implied side-effect of the proposed strategy renders a more balanced distribution of street space for all cities under study.

We foresee that sidewalk networks will trigger much needed research on a wealth of topics that precede, and will no doubt extend beyond, the current health crisis, starting with the general lack of available public data on sidewalk infrastructure. By now, cities will have to consider whether street pacification strategies belong to the set of extraordinary --and thus temporary-- measures to flatten the contagion curve and associated socioeconomic effects of the COVID-19, or if such strategies should be counted among other \cite{xu2020deconstructing, batty2020coronavirus} long-term actions towards the end of car-dominated cities.

\section*{Acknowledgments}
D.R., A.S-R. and J.B-H. acknowledge financial support from the Direcci\'on General de Tr\'afico (Spain), Project No. SPIP2017-02263. DR acknowledges as well the support of a doctoral grant from the Universitat Oberta de Catalunya (UOC). This work was supported by the Berkeley DeepDrive (BDD) and the University of California Institute of Transportation Studies (UC ITS) research grants.


\begin{thebibliography}{40}%
\makeatletter
\providecommand \@ifxundefined [1]{%
 \@ifx{#1\undefined}
}%
\providecommand \@ifnum [1]{%
 \ifnum #1\expandafter \@firstoftwo
 \else \expandafter \@secondoftwo
 \fi
}%
\providecommand \@ifx [1]{%
 \ifx #1\expandafter \@firstoftwo
 \else \expandafter \@secondoftwo
 \fi
}%
\providecommand \natexlab [1]{#1}%
\providecommand \enquote  [1]{``#1''}%
\providecommand \bibnamefont  [1]{#1}%
\providecommand \bibfnamefont [1]{#1}%
\providecommand \citenamefont [1]{#1}%
\providecommand \href@noop [0]{\@secondoftwo}%
\providecommand \href [0]{\begingroup \@sanitize@url \@href}%
\providecommand \@href[1]{\@@startlink{#1}\@@href}%
\providecommand \@@href[1]{\endgroup#1\@@endlink}%
\providecommand \@sanitize@url [0]{\catcode `\\12\catcode `\$12\catcode
  `\&12\catcode `\#12\catcode `\^12\catcode `\_12\catcode `\%12\relax}%
\providecommand \@@startlink[1]{}%
\providecommand \@@endlink[0]{}%
\providecommand \url  [0]{\begingroup\@sanitize@url \@url }%
\providecommand \@url [1]{\endgroup\@href {#1}{\urlprefix }}%
\providecommand \urlprefix  [0]{URL }%
\providecommand \Eprint [0]{\href }%
\providecommand \doibase [0]{http://dx.doi.org/}%
\providecommand \selectlanguage [0]{\@gobble}%
\providecommand \bibinfo  [0]{\@secondoftwo}%
\providecommand \bibfield  [0]{\@secondoftwo}%
\providecommand \translation [1]{[#1]}%
\providecommand \BibitemOpen [0]{}%
\providecommand \bibitemStop [0]{}%
\providecommand \bibitemNoStop [0]{.\EOS\space}%
\providecommand \EOS [0]{\spacefactor3000\relax}%
\providecommand \BibitemShut  [1]{\csname bibitem#1\endcsname}%
\let\auto@bib@innerbib\@empty
\bibitem [{\citenamefont {Gordon}(2020)}]{gordon2020}%
  \BibitemOpen
  \bibfield  {author} {\bibinfo {author} {\bibfnamefont {A.}~\bibnamefont
  {Gordon}},\ }\href
  {https://www.vice.com/en_us/article/k7evnv/trying-to-maintain-social-distancing-on-new-york-city-sidewalks-good-luck}
  {\enquote {\bibinfo {title} {Trying to maintain social distancing on new york
  city sidewalks? good luck},}\ } (\bibinfo {year} {2020})\BibitemShut
  {NoStop}%
\bibitem [{\citenamefont {Fleischer}(2020)}]{fleischer2020}%
  \BibitemOpen
  \bibfield  {author} {\bibinfo {author} {\bibfnamefont {M.}~\bibnamefont
  {Fleischer}},\ }\href
  {https://www.latimes.com/opinion/story/2020-04-02/sidewalks-coronavirus-contagion-pedestrians}
  {\enquote {\bibinfo {title} {Opinion: Crowded sidewalks are vectors for
  coronavirus infection. take streets back from cars and let people use
  them},}\ } (\bibinfo {year} {2020})\BibitemShut {NoStop}%
\bibitem [{\citenamefont {Sadik-Khan}(2020)}]{sadikkhan2020}%
  \BibitemOpen
  \bibfield  {author} {\bibinfo {author} {\bibfnamefont {J.}~\bibnamefont
  {Sadik-Khan}},\ }\href
  {https://www.theguardian.com/cities/2020/sep/04/janette-sadik-khan-we-must-rethink-our-streets-to-create-the-six-foot-city}
  {\enquote {\bibinfo {title} {Janette sadik-khan: We must rethink our streets
  to create the six-foot city},}\ } (\bibinfo {year} {2020})\BibitemShut
  {NoStop}%
\bibitem [{\citenamefont {Block}(2020)}]{india2020}%
  \BibitemOpen
  \bibfield  {author} {\bibinfo {author} {\bibfnamefont {I.}~\bibnamefont
  {Block}},\ }\href
  {https://www.dezeen.com/2020/05/07/london-new-york-paris-milan-cyclists-pedestrians/}
  {\enquote {\bibinfo {title} {London, new york, paris and milan give streets
  to cyclists and pedestrians},}\ } (\bibinfo {year} {2020})\BibitemShut
  {NoStop}%
\bibitem [{\citenamefont {Guerrero}(2020)}]{guerrero2020}%
  \BibitemOpen
  \bibfield  {author} {\bibinfo {author} {\bibfnamefont {D.}~\bibnamefont
  {Guerrero}},\ }\href
  {https://www.lavanguardia.com/local/barcelona/20200425/48704357907/barcelona-espacio-coches-aceras-carriles-bici-desconfinamiento-coronavirus.html}
  {\enquote {\bibinfo {title} {Barcelona convertir\'a espacios destinados al
  coche en aceras y carriles bici},}\ } (\bibinfo {year} {2020})\BibitemShut
  {NoStop}%
\bibitem [{\citenamefont {Colville-Andersen}(2018)}]{arrogance2018}%
  \BibitemOpen
  \bibfield  {author} {\bibinfo {author} {\bibfnamefont {M.}~\bibnamefont
  {Colville-Andersen}},\ }\enquote {\bibinfo {title} {Copenhagenize},}\ \
  (\bibinfo  {publisher} {Island Press},\ \bibinfo {address} {Washington, DC},\
  \bibinfo {year} {2018})\ Chap.\ \bibinfo {chapter} {The Arrogance of
  Space}\BibitemShut {NoStop}%
\bibitem [{\citenamefont {{Creutzig}}\ \emph {et~al.}(2020)\citenamefont
  {{Creutzig}}, \citenamefont {{Javaid}}, \citenamefont {{Soomauroo}},
  \citenamefont {{Lohrey}}, \citenamefont {{Milojevic-Dupont}}, \citenamefont
  {{Ramakrishnan}}, \citenamefont {{Sethi}}, \citenamefont {{Liu}},
  \citenamefont {{Niamir}}, \citenamefont {{d'Amour}}, \citenamefont
  {{Weddige}}, \citenamefont {{Lenzi}}, \citenamefont {{Kowarsch}},
  \citenamefont {{Arndt}}, \citenamefont {{Baumann}}, \citenamefont
  {{Betzien}}, \citenamefont {{Fonkwa}}, \citenamefont {{Huber}}, \citenamefont
  {{Mendez}}, \citenamefont {{Misiou}}, \citenamefont {{Pearce}}, \citenamefont
  {{Radman}}, \citenamefont {{Skaloud}},\ and\ \citenamefont
  {{Zausch}}}]{creutzig2020fair}%
  \BibitemOpen
  \bibfield  {author} {\bibinfo {author} {\bibfnamefont {F.}~\bibnamefont
  {{Creutzig}}}, \bibinfo {author} {\bibfnamefont {A.}~\bibnamefont
  {{Javaid}}}, \bibinfo {author} {\bibfnamefont {Z.}~\bibnamefont
  {{Soomauroo}}}, \bibinfo {author} {\bibfnamefont {S.}~\bibnamefont
  {{Lohrey}}}, \bibinfo {author} {\bibfnamefont {N.}~\bibnamefont
  {{Milojevic-Dupont}}}, \bibinfo {author} {\bibfnamefont {A.}~\bibnamefont
  {{Ramakrishnan}}}, \bibinfo {author} {\bibfnamefont {M.}~\bibnamefont
  {{Sethi}}}, \bibinfo {author} {\bibfnamefont {L.}~\bibnamefont {{Liu}}},
  \bibinfo {author} {\bibfnamefont {L.}~\bibnamefont {{Niamir}}}, \bibinfo
  {author} {\bibfnamefont {C.~B.}\ \bibnamefont {{d'Amour}}}, \bibinfo {author}
  {\bibfnamefont {U.}~\bibnamefont {{Weddige}}}, \bibinfo {author}
  {\bibfnamefont {D.}~\bibnamefont {{Lenzi}}}, \bibinfo {author} {\bibfnamefont
  {M.}~\bibnamefont {{Kowarsch}}}, \bibinfo {author} {\bibfnamefont
  {L.}~\bibnamefont {{Arndt}}}, \bibinfo {author} {\bibfnamefont
  {L.}~\bibnamefont {{Baumann}}}, \bibinfo {author} {\bibfnamefont
  {J.}~\bibnamefont {{Betzien}}}, \bibinfo {author} {\bibfnamefont
  {L.}~\bibnamefont {{Fonkwa}}}, \bibinfo {author} {\bibfnamefont
  {B.}~\bibnamefont {{Huber}}}, \bibinfo {author} {\bibfnamefont
  {E.}~\bibnamefont {{Mendez}}}, \bibinfo {author} {\bibfnamefont
  {A.}~\bibnamefont {{Misiou}}}, \bibinfo {author} {\bibfnamefont
  {C.}~\bibnamefont {{Pearce}}}, \bibinfo {author} {\bibfnamefont
  {P.}~\bibnamefont {{Radman}}}, \bibinfo {author} {\bibfnamefont
  {P.}~\bibnamefont {{Skaloud}}}, \ and\ \bibinfo {author} {\bibfnamefont
  {J.~M.}\ \bibnamefont {{Zausch}}},\ }\href@noop {} {\bibfield  {journal}
  {\bibinfo  {journal} {Transport Reviews}\ ,\ \bibinfo {pages} {1}} (\bibinfo
  {year} {2020})}\BibitemShut {NoStop}%
\bibitem [{\citenamefont {Batty}(2012)}]{Batty2012}%
  \BibitemOpen
  \bibfield  {author} {\bibinfo {author} {\bibfnamefont {M.}~\bibnamefont
  {Batty}},\ }\href {\doibase https://doi.org/10.1016/j.cities.2011.11.008}
  {\bibfield  {journal} {\bibinfo  {journal} {Cities}\ }\textbf {\bibinfo
  {volume} {29}},\ \bibinfo {pages} {S9 } (\bibinfo {year} {2012})},\ \bibinfo
  {note} {current Research on Cities}\BibitemShut {NoStop}%
\bibitem [{\citenamefont {Barthelemy}(2018)}]{barthelemy2018morphogenesis}%
  \BibitemOpen
  \bibfield  {author} {\bibinfo {author} {\bibfnamefont {M.}~\bibnamefont
  {Barthelemy}},\ }\href@noop {} {\emph {\bibinfo {title} {Morphogenesis of
  spatial networks}}}\ (\bibinfo  {publisher} {Springer},\ \bibinfo {year}
  {2018})\BibitemShut {NoStop}%
\bibitem [{\citenamefont {Radicchi}(2015)}]{radicchi2015percolation}%
  \BibitemOpen
  \bibfield  {author} {\bibinfo {author} {\bibfnamefont {F.}~\bibnamefont
  {Radicchi}},\ }\href@noop {} {\bibfield  {journal} {\bibinfo  {journal}
  {Nature Physics}\ }\textbf {\bibinfo {volume} {11}},\ \bibinfo {pages} {597}
  (\bibinfo {year} {2015})}\BibitemShut {NoStop}%
\bibitem [{\citenamefont {Graham}(2020)}]{Graham2020}%
  \BibitemOpen
  \bibfield  {author} {\bibinfo {author} {\bibfnamefont {B.~S.}\ \bibnamefont
  {Graham}},\ }\href@noop {} {\bibfield  {journal} {\bibinfo  {journal}
  {Science}\ }\textbf {\bibinfo {volume} {368}},\ \bibinfo {pages} {945}
  (\bibinfo {year} {2020})}\BibitemShut {NoStop}%
\bibitem [{\citenamefont {Thorp}(2020)}]{Thorp2020}%
  \BibitemOpen
  \bibfield  {author} {\bibinfo {author} {\bibfnamefont {H.~H.}\ \bibnamefont
  {Thorp}},\ }\href@noop {} {\bibfield  {journal} {\bibinfo  {journal}
  {Science}\ }\textbf {\bibinfo {volume} {369}},\ \bibinfo {pages} {885}
  (\bibinfo {year} {2020})}\BibitemShut {NoStop}%
\bibitem [{\citenamefont {Le}\ \emph {et~al.}(2020)\citenamefont {Le},
  \citenamefont {Andreadakis}, \citenamefont {Kumar}, \citenamefont {Roman},
  \citenamefont {Tollefsen}, \citenamefont {Saville},\ and\ \citenamefont
  {Mayhew}}]{le2020covid}%
  \BibitemOpen
  \bibfield  {author} {\bibinfo {author} {\bibfnamefont {T.~T.}\ \bibnamefont
  {Le}}, \bibinfo {author} {\bibfnamefont {Z.}~\bibnamefont {Andreadakis}},
  \bibinfo {author} {\bibfnamefont {A.}~\bibnamefont {Kumar}}, \bibinfo
  {author} {\bibfnamefont {R.~G.}\ \bibnamefont {Roman}}, \bibinfo {author}
  {\bibfnamefont {S.}~\bibnamefont {Tollefsen}}, \bibinfo {author}
  {\bibfnamefont {M.}~\bibnamefont {Saville}}, \ and\ \bibinfo {author}
  {\bibfnamefont {S.}~\bibnamefont {Mayhew}},\ }\href@noop {} {\bibfield
  {journal} {\bibinfo  {journal} {Nat Rev Drug Discov}\ }\textbf {\bibinfo
  {volume} {19}},\ \bibinfo {pages} {305} (\bibinfo {year} {2020})}\BibitemShut
  {NoStop}%
\bibitem [{\citenamefont {Callaway}(2020)}]{callaway2020coronavirus}%
  \BibitemOpen
  \bibfield  {author} {\bibinfo {author} {\bibfnamefont {E.}~\bibnamefont
  {Callaway}},\ }\href@noop {} {\bibfield  {journal} {\bibinfo  {journal}
  {Nature}\ }\textbf {\bibinfo {volume} {583}},\ \bibinfo {pages} {669}
  (\bibinfo {year} {2020})}\BibitemShut {NoStop}%
\bibitem [{\citenamefont {Mossong}\ \emph {et~al.}(2008)\citenamefont
  {Mossong}, \citenamefont {Hens}, \citenamefont {Jit}, \citenamefont
  {Beutels}, \citenamefont {Auranen}, \citenamefont {Mikolajczyk},
  \citenamefont {Massari}, \citenamefont {Salmaso}, \citenamefont {Tomba},
  \citenamefont {Wallinga}, \citenamefont {Heijne}, \citenamefont
  {Sadkowska-Todys}, \citenamefont {Rosinska},\ and\ \citenamefont
  {Edmunds}}]{Mossong_2008}%
  \BibitemOpen
  \bibfield  {author} {\bibinfo {author} {\bibfnamefont {J.}~\bibnamefont
  {Mossong}}, \bibinfo {author} {\bibfnamefont {N.}~\bibnamefont {Hens}},
  \bibinfo {author} {\bibfnamefont {M.}~\bibnamefont {Jit}}, \bibinfo {author}
  {\bibfnamefont {P.}~\bibnamefont {Beutels}}, \bibinfo {author} {\bibfnamefont
  {K.}~\bibnamefont {Auranen}}, \bibinfo {author} {\bibfnamefont
  {R.}~\bibnamefont {Mikolajczyk}}, \bibinfo {author} {\bibfnamefont
  {M.}~\bibnamefont {Massari}}, \bibinfo {author} {\bibfnamefont
  {S.}~\bibnamefont {Salmaso}}, \bibinfo {author} {\bibfnamefont {G.~S.}\
  \bibnamefont {Tomba}}, \bibinfo {author} {\bibfnamefont {J.}~\bibnamefont
  {Wallinga}}, \bibinfo {author} {\bibfnamefont {J.}~\bibnamefont {Heijne}},
  \bibinfo {author} {\bibfnamefont {M.}~\bibnamefont {Sadkowska-Todys}},
  \bibinfo {author} {\bibfnamefont {M.}~\bibnamefont {Rosinska}}, \ and\
  \bibinfo {author} {\bibfnamefont {W.~J.}\ \bibnamefont {Edmunds}},\ }\href
  {\doibase 10.1371/journal.pmed.0050074} {\bibfield  {journal} {\bibinfo
  {journal} {{PLoS} Medicine}\ }\textbf {\bibinfo {volume} {5}},\ \bibinfo
  {pages} {e74} (\bibinfo {year} {2008})}\BibitemShut {NoStop}%
\bibitem [{\citenamefont {Hsiang}\ \emph {et~al.}(2020)\citenamefont {Hsiang},
  \citenamefont {Allen}, \citenamefont {Annan-Phan}, \citenamefont {Bell},
  \citenamefont {Bolliger}, \citenamefont {Chong}, \citenamefont
  {Druckenmiller}, \citenamefont {Huang}, \citenamefont {Hultgren},
  \citenamefont {Krasovich}, \citenamefont {Lau}, \citenamefont {Lee},
  \citenamefont {Rolf}, \citenamefont {Tseng},\ and\ \citenamefont
  {Wu}}]{hsiang2020}%
  \BibitemOpen
  \bibfield  {author} {\bibinfo {author} {\bibfnamefont {S.}~\bibnamefont
  {Hsiang}}, \bibinfo {author} {\bibfnamefont {D.}~\bibnamefont {Allen}},
  \bibinfo {author} {\bibfnamefont {S.}~\bibnamefont {Annan-Phan}}, \bibinfo
  {author} {\bibfnamefont {K.}~\bibnamefont {Bell}}, \bibinfo {author}
  {\bibfnamefont {I.}~\bibnamefont {Bolliger}}, \bibinfo {author}
  {\bibfnamefont {T.}~\bibnamefont {Chong}}, \bibinfo {author} {\bibfnamefont
  {H.}~\bibnamefont {Druckenmiller}}, \bibinfo {author} {\bibfnamefont {L.~Y.}\
  \bibnamefont {Huang}}, \bibinfo {author} {\bibfnamefont {A.}~\bibnamefont
  {Hultgren}}, \bibinfo {author} {\bibfnamefont {E.}~\bibnamefont {Krasovich}},
  \bibinfo {author} {\bibfnamefont {P.}~\bibnamefont {Lau}}, \bibinfo {author}
  {\bibfnamefont {J.}~\bibnamefont {Lee}}, \bibinfo {author} {\bibfnamefont
  {E.}~\bibnamefont {Rolf}}, \bibinfo {author} {\bibfnamefont {J.}~\bibnamefont
  {Tseng}}, \ and\ \bibinfo {author} {\bibfnamefont {T.}~\bibnamefont {Wu}},\
  }\href {\doibase 10.1038/s41586-020-2404-8} {\bibfield  {journal} {\bibinfo
  {journal} {Nature}\ }\textbf {\bibinfo {volume} {584}},\ \bibinfo {pages}
  {262} (\bibinfo {year} {2020})}\BibitemShut {NoStop}%
\bibitem [{\citenamefont {Chu}\ \emph {et~al.}(2020)\citenamefont {Chu},
  \citenamefont {Akl}, \citenamefont {Duda}, \citenamefont {Solo},
  \citenamefont {Yaacoub}, \citenamefont {Sch{\"u}nemann}, \citenamefont
  {El-harakeh}, \citenamefont {Bognanni}, \citenamefont {Lotfi}, \citenamefont
  {Loeb} \emph {et~al.}}]{chu2020physical}%
  \BibitemOpen
  \bibfield  {author} {\bibinfo {author} {\bibfnamefont {D.~K.}\ \bibnamefont
  {Chu}}, \bibinfo {author} {\bibfnamefont {E.~A.}\ \bibnamefont {Akl}},
  \bibinfo {author} {\bibfnamefont {S.}~\bibnamefont {Duda}}, \bibinfo {author}
  {\bibfnamefont {K.}~\bibnamefont {Solo}}, \bibinfo {author} {\bibfnamefont
  {S.}~\bibnamefont {Yaacoub}}, \bibinfo {author} {\bibfnamefont {H.~J.}\
  \bibnamefont {Sch{\"u}nemann}}, \bibinfo {author} {\bibfnamefont
  {A.}~\bibnamefont {El-harakeh}}, \bibinfo {author} {\bibfnamefont
  {A.}~\bibnamefont {Bognanni}}, \bibinfo {author} {\bibfnamefont
  {T.}~\bibnamefont {Lotfi}}, \bibinfo {author} {\bibfnamefont
  {M.}~\bibnamefont {Loeb}},  \emph {et~al.},\ }\href@noop {} {\bibfield
  {journal} {\bibinfo  {journal} {The Lancet}\ } (\bibinfo {year}
  {2020})}\BibitemShut {NoStop}%
\bibitem [{WHO(2020)}]{WHO}%
  \BibitemOpen
  \href@noop {} {\emph {\bibinfo {title} {Moving around during the COVID-19
  outbreak}}},\ \bibinfo {type} {Tech. Rep.}\ (\bibinfo  {institution} {World
  Health Organization},\ \bibinfo {year} {2020})\BibitemShut {NoStop}%
\bibitem [{\citenamefont {Sadik-Khan}()}]{streets2020Nacto}%
  \BibitemOpen
  \bibfield  {author} {\bibinfo {author} {\bibfnamefont {J.}~\bibnamefont
  {Sadik-Khan}},\ }\href@noop {} {\enquote {\bibinfo {title} {Streets for
  pandemic response and recovery},}\ }\bibinfo {note} {Accessed:
  7-8-2020}\BibitemShut {NoStop}%
\bibitem [{\citenamefont {Troko}\ \emph {et~al.}(2011)\citenamefont {Troko},
  \citenamefont {Myles}, \citenamefont {Gibson}, \citenamefont {Hashim},
  \citenamefont {Enstone}, \citenamefont {Kingdon}, \citenamefont {Packham},
  \citenamefont {Amin}, \citenamefont {Hayward},\ and\ \citenamefont
  {Van-Tam}}]{troko2011public}%
  \BibitemOpen
  \bibfield  {author} {\bibinfo {author} {\bibfnamefont {J.}~\bibnamefont
  {Troko}}, \bibinfo {author} {\bibfnamefont {P.}~\bibnamefont {Myles}},
  \bibinfo {author} {\bibfnamefont {J.}~\bibnamefont {Gibson}}, \bibinfo
  {author} {\bibfnamefont {A.}~\bibnamefont {Hashim}}, \bibinfo {author}
  {\bibfnamefont {J.}~\bibnamefont {Enstone}}, \bibinfo {author} {\bibfnamefont
  {S.}~\bibnamefont {Kingdon}}, \bibinfo {author} {\bibfnamefont
  {C.}~\bibnamefont {Packham}}, \bibinfo {author} {\bibfnamefont
  {S.}~\bibnamefont {Amin}}, \bibinfo {author} {\bibfnamefont {A.}~\bibnamefont
  {Hayward}}, \ and\ \bibinfo {author} {\bibfnamefont {J.~N.}\ \bibnamefont
  {Van-Tam}},\ }\href@noop {} {\bibfield  {journal} {\bibinfo  {journal} {BMC
  infectious diseases}\ }\textbf {\bibinfo {volume} {11}},\ \bibinfo {pages}
  {1} (\bibinfo {year} {2011})}\BibitemShut {NoStop}%
\bibitem [{\citenamefont {Goldbaum}(2020)}]{goldbaum2020}%
  \BibitemOpen
  \bibfield  {author} {\bibinfo {author} {\bibfnamefont {C.}~\bibnamefont
  {Goldbaum}},\ }\href
  {https://www.nytimes.com/2020/03/24/nyregion/coronavirus-nyc-mta-cuts-.html}
  {\enquote {\bibinfo {title} {Subway service is cut by a quarter because of
  coronavirus},}\ } (\bibinfo {year} {2020})\BibitemShut {NoStop}%
\bibitem [{\citenamefont {Batty}(2020)}]{batty2020coronavirus}%
  \BibitemOpen
  \bibfield  {author} {\bibinfo {author} {\bibfnamefont {M.}~\bibnamefont
  {Batty}},\ }\href@noop {} {\bibfield  {journal} {\bibinfo  {journal}
  {Environment and Planning B: Urban Analytics and City Science}\ }\textbf
  {\bibinfo {volume} {47}},\ \bibinfo {pages} {547} (\bibinfo {year}
  {2020})}\BibitemShut {NoStop}%
\bibitem [{\citenamefont {Wu}\ \emph {et~al.}(2020)\citenamefont {Wu},
  \citenamefont {Nethery}, \citenamefont {Sabath}, \citenamefont {Braun},\ and\
  \citenamefont {Dominici}}]{wu2020exposure}%
  \BibitemOpen
  \bibfield  {author} {\bibinfo {author} {\bibfnamefont {X.}~\bibnamefont
  {Wu}}, \bibinfo {author} {\bibfnamefont {R.~C.}\ \bibnamefont {Nethery}},
  \bibinfo {author} {\bibfnamefont {B.~M.}\ \bibnamefont {Sabath}}, \bibinfo
  {author} {\bibfnamefont {D.}~\bibnamefont {Braun}}, \ and\ \bibinfo {author}
  {\bibfnamefont {F.}~\bibnamefont {Dominici}},\ }\href@noop {} {\bibfield
  {journal} {\bibinfo  {journal} {medRxiv}\ } (\bibinfo {year}
  {2020})}\BibitemShut {NoStop}%
\bibitem [{\citenamefont {Stier}\ \emph {et~al.}(2020)\citenamefont {Stier},
  \citenamefont {Berman},\ and\ \citenamefont {Bettencourt}}]{stier2020covid}%
  \BibitemOpen
  \bibfield  {author} {\bibinfo {author} {\bibfnamefont {A.}~\bibnamefont
  {Stier}}, \bibinfo {author} {\bibfnamefont {M.}~\bibnamefont {Berman}}, \
  and\ \bibinfo {author} {\bibfnamefont {L.}~\bibnamefont {Bettencourt}},\
  }\href@noop {} {\bibfield  {journal} {\bibinfo  {journal} {Mansueto Institute
  for Urban Innovation Research Paper Forthcoming}\ } (\bibinfo {year}
  {2020})}\BibitemShut {NoStop}%
\bibitem [{\citenamefont {De~Vos}(2020)}]{deVos2020}%
  \BibitemOpen
  \bibfield  {author} {\bibinfo {author} {\bibfnamefont {J.}~\bibnamefont
  {De~Vos}},\ }\href@noop {} {\bibfield  {journal} {\bibinfo  {journal}
  {Transportation Research Interdisciplinary Perspectives}\ ,\ \bibinfo {pages}
  {100121}} (\bibinfo {year} {2020})}\BibitemShut {NoStop}%
\bibitem [{\citenamefont {Seale}\ \emph {et~al.}(2020)\citenamefont {Seale},
  \citenamefont {Heywood}, \citenamefont {Leask}, \citenamefont {Steel},
  \citenamefont {Thomas}, \citenamefont {Durrheim}, \citenamefont {Bolsewicz},\
  and\ \citenamefont {Kaur}}]{seale2020covid}%
  \BibitemOpen
  \bibfield  {author} {\bibinfo {author} {\bibfnamefont {H.}~\bibnamefont
  {Seale}}, \bibinfo {author} {\bibfnamefont {A.~E.}\ \bibnamefont {Heywood}},
  \bibinfo {author} {\bibfnamefont {J.}~\bibnamefont {Leask}}, \bibinfo
  {author} {\bibfnamefont {M.}~\bibnamefont {Steel}}, \bibinfo {author}
  {\bibfnamefont {S.}~\bibnamefont {Thomas}}, \bibinfo {author} {\bibfnamefont
  {D.~N.}\ \bibnamefont {Durrheim}}, \bibinfo {author} {\bibfnamefont
  {K.}~\bibnamefont {Bolsewicz}}, \ and\ \bibinfo {author} {\bibfnamefont
  {R.}~\bibnamefont {Kaur}},\ }\href@noop {} {\bibfield  {journal} {\bibinfo
  {journal} {medRxiv}\ } (\bibinfo {year} {2020})}\BibitemShut {NoStop}%
\bibitem [{\citenamefont {M{\ae}kel{\ae}}\ \emph {et~al.}(2020)\citenamefont
  {M{\ae}kel{\ae}}, \citenamefont {Reggev}, \citenamefont {Dutra},
  \citenamefont {Tamayo}, \citenamefont {Silva-Sobrinho}, \citenamefont
  {Klevjer},\ and\ \citenamefont {Pfuhl}}]{Mkel2020}%
  \BibitemOpen
  \bibfield  {author} {\bibinfo {author} {\bibfnamefont {M.~J.}\ \bibnamefont
  {M{\ae}kel{\ae}}}, \bibinfo {author} {\bibfnamefont {N.}~\bibnamefont
  {Reggev}}, \bibinfo {author} {\bibfnamefont {N.}~\bibnamefont {Dutra}},
  \bibinfo {author} {\bibfnamefont {R.~M.}\ \bibnamefont {Tamayo}}, \bibinfo
  {author} {\bibfnamefont {R.~A.}\ \bibnamefont {Silva-Sobrinho}}, \bibinfo
  {author} {\bibfnamefont {K.}~\bibnamefont {Klevjer}}, \ and\ \bibinfo
  {author} {\bibfnamefont {G.}~\bibnamefont {Pfuhl}},\ }\href {\doibase
  10.1098/rsos.200644} {\bibfield  {journal} {\bibinfo  {journal} {Royal
  Society Open Science}\ }\textbf {\bibinfo {volume} {7}},\ \bibinfo {pages}
  {200644} (\bibinfo {year} {2020})}\BibitemShut {NoStop}%
\bibitem [{\citenamefont {NACTO}(2013)}]{nacto2013urban}%
  \BibitemOpen
  \bibfield  {author} {\bibinfo {author} {\bibnamefont {NACTO}},\ }\href@noop
  {} {\emph {\bibinfo {title} {Urban street design guide}}}\ (\bibinfo
  {publisher} {Island Press/Center for Resource Economics Washington, DC},\
  \bibinfo {year} {2013})\BibitemShut {NoStop}%
\bibitem [{\citenamefont {Albert}\ \emph {et~al.}(2000)\citenamefont {Albert},
  \citenamefont {Jeong},\ and\ \citenamefont {Barab{\'a}si}}]{albert2000error}%
  \BibitemOpen
  \bibfield  {author} {\bibinfo {author} {\bibfnamefont {R.}~\bibnamefont
  {Albert}}, \bibinfo {author} {\bibfnamefont {H.}~\bibnamefont {Jeong}}, \
  and\ \bibinfo {author} {\bibfnamefont {A.-L.}\ \bibnamefont {Barab{\'a}si}},\
  }\href@noop {} {\bibfield  {journal} {\bibinfo  {journal} {Nature}\ }\textbf
  {\bibinfo {volume} {406}},\ \bibinfo {pages} {378} (\bibinfo {year}
  {2000})}\BibitemShut {NoStop}%
\bibitem [{\citenamefont {Cohen}\ \emph {et~al.}(2001)\citenamefont {Cohen},
  \citenamefont {Erez}, \citenamefont {Ben-Avraham},\ and\ \citenamefont
  {Havlin}}]{cohen2001breakdown}%
  \BibitemOpen
  \bibfield  {author} {\bibinfo {author} {\bibfnamefont {R.}~\bibnamefont
  {Cohen}}, \bibinfo {author} {\bibfnamefont {K.}~\bibnamefont {Erez}},
  \bibinfo {author} {\bibfnamefont {D.}~\bibnamefont {Ben-Avraham}}, \ and\
  \bibinfo {author} {\bibfnamefont {S.}~\bibnamefont {Havlin}},\ }\href@noop {}
  {\bibfield  {journal} {\bibinfo  {journal} {Physical Review Letters}\
  }\textbf {\bibinfo {volume} {86}},\ \bibinfo {pages} {3682} (\bibinfo {year}
  {2001})}\BibitemShut {NoStop}%
\bibitem [{\citenamefont {Abbar}\ \emph {et~al.}(2018)\citenamefont {Abbar},
  \citenamefont {Zanouda},\ and\ \citenamefont
  {Borge-Holthoefer}}]{abbar2018structural}%
  \BibitemOpen
  \bibfield  {author} {\bibinfo {author} {\bibfnamefont {S.}~\bibnamefont
  {Abbar}}, \bibinfo {author} {\bibfnamefont {T.}~\bibnamefont {Zanouda}}, \
  and\ \bibinfo {author} {\bibfnamefont {J.}~\bibnamefont {Borge-Holthoefer}},\
  }\href@noop {} {\bibfield  {journal} {\bibinfo  {journal} {Data Mining and
  Knowledge Discovery}\ }\textbf {\bibinfo {volume} {32}},\ \bibinfo {pages}
  {830} (\bibinfo {year} {2018})}\BibitemShut {NoStop}%
\bibitem [{\citenamefont {Janson}\ \emph {et~al.}(1993)\citenamefont {Janson},
  \citenamefont {Knuth}, \citenamefont {{\L}uczak},\ and\ \citenamefont
  {Pittel}}]{janson1993birth}%
  \BibitemOpen
  \bibfield  {author} {\bibinfo {author} {\bibfnamefont {S.}~\bibnamefont
  {Janson}}, \bibinfo {author} {\bibfnamefont {D.~E.}\ \bibnamefont {Knuth}},
  \bibinfo {author} {\bibfnamefont {T.}~\bibnamefont {{\L}uczak}}, \ and\
  \bibinfo {author} {\bibfnamefont {B.}~\bibnamefont {Pittel}},\ }\href@noop {}
  {\bibfield  {journal} {\bibinfo  {journal} {Random Structures \& Algorithms}\
  }\textbf {\bibinfo {volume} {4}},\ \bibinfo {pages} {233} (\bibinfo {year}
  {1993})}\BibitemShut {NoStop}%
\bibitem [{\citenamefont {Molloy}\ and\ \citenamefont
  {Reed}(1998)}]{molloy1998size}%
  \BibitemOpen
  \bibfield  {author} {\bibinfo {author} {\bibfnamefont {M.}~\bibnamefont
  {Molloy}}\ and\ \bibinfo {author} {\bibfnamefont {B.}~\bibnamefont {Reed}},\
  }\href@noop {} {\bibfield  {journal} {\bibinfo  {journal} {Combinatorics
  probability and computing}\ }\textbf {\bibinfo {volume} {7}},\ \bibinfo
  {pages} {295} (\bibinfo {year} {1998})}\BibitemShut {NoStop}%
\bibitem [{\citenamefont {Brindle}(1991)}]{brindle1991traffic}%
  \BibitemOpen
  \bibfield  {author} {\bibinfo {author} {\bibfnamefont {B.}~\bibnamefont
  {Brindle}},\ }\href@noop {} {\bibfield  {journal} {\bibinfo  {journal}
  {Australian Road Research}\ }\textbf {\bibinfo {volume} {21}},\ \bibinfo
  {pages} {37} (\bibinfo {year} {1991})}\BibitemShut {NoStop}%
\bibitem [{hab(2016)}]{habitat2016united}%
  \BibitemOpen
  \href@noop {} {\emph {\bibinfo {title} {United Nations Conference on Housing
  and Sustainable Urban Development (UN-HABITAT III)}}}\ (\bibinfo {year}
  {2016})\BibitemShut {NoStop}%
\bibitem [{\citenamefont {Karndacharuk}\ \emph {et~al.}(2014)\citenamefont
  {Karndacharuk}, \citenamefont {Wilson},\ and\ \citenamefont
  {Dunn}}]{Karndacharuk_2014}%
  \BibitemOpen
  \bibfield  {author} {\bibinfo {author} {\bibfnamefont {A.}~\bibnamefont
  {Karndacharuk}}, \bibinfo {author} {\bibfnamefont {D.~J.}\ \bibnamefont
  {Wilson}}, \ and\ \bibinfo {author} {\bibfnamefont {R.}~\bibnamefont
  {Dunn}},\ }\href {\doibase 10.1080/01441647.2014.893038} {\bibfield
  {journal} {\bibinfo  {journal} {Transport Reviews}\ }\textbf {\bibinfo
  {volume} {34}},\ \bibinfo {pages} {190} (\bibinfo {year} {2014})}\BibitemShut
  {NoStop}%
\bibitem [{\citenamefont {Guimer\`a}\ \emph {et~al.}(2002)\citenamefont
  {Guimer\`a}, \citenamefont {D\'{\i}az-Guilera}, \citenamefont {Vega-Redondo},
  \citenamefont {Cabrales},\ and\ \citenamefont {Arenas}}]{Guimera2002}%
  \BibitemOpen
  \bibfield  {author} {\bibinfo {author} {\bibfnamefont {R.}~\bibnamefont
  {Guimer\`a}}, \bibinfo {author} {\bibfnamefont {A.}~\bibnamefont
  {D\'{\i}az-Guilera}}, \bibinfo {author} {\bibfnamefont {F.}~\bibnamefont
  {Vega-Redondo}}, \bibinfo {author} {\bibfnamefont {A.}~\bibnamefont
  {Cabrales}}, \ and\ \bibinfo {author} {\bibfnamefont {A.}~\bibnamefont
  {Arenas}},\ }\href {\doibase https://doi.org/10.1103/PhysRevLett.89.248701}
  {\bibfield  {journal} {\bibinfo  {journal} {Phys. Rev. Lett.}\ }\textbf
  {\bibinfo {volume} {89}},\ \bibinfo {pages} {248701} (\bibinfo {year}
  {2002})}\BibitemShut {NoStop}%
\bibitem [{\citenamefont {Sol{\'e}-Ribalta}\ \emph {et~al.}(2018)\citenamefont
  {Sol{\'e}-Ribalta}, \citenamefont {G{\'o}mez},\ and\ \citenamefont
  {Arenas}}]{sole2018decongestion}%
  \BibitemOpen
  \bibfield  {author} {\bibinfo {author} {\bibfnamefont {A.}~\bibnamefont
  {Sol{\'e}-Ribalta}}, \bibinfo {author} {\bibfnamefont {S.}~\bibnamefont
  {G{\'o}mez}}, \ and\ \bibinfo {author} {\bibfnamefont {A.}~\bibnamefont
  {Arenas}},\ }\href@noop {} {\bibfield  {journal} {\bibinfo  {journal}
  {Networks and Spatial Economics}\ }\textbf {\bibinfo {volume} {18}},\
  \bibinfo {pages} {33} (\bibinfo {year} {2018})}\BibitemShut {NoStop}%
\bibitem [{\citenamefont {Sol{\'e}-Ribalta}\ \emph {et~al.}(2019)\citenamefont
  {Sol{\'e}-Ribalta}, \citenamefont {Arenas},\ and\ \citenamefont
  {G{\'o}mez}}]{sole2019effect}%
  \BibitemOpen
  \bibfield  {author} {\bibinfo {author} {\bibfnamefont {A.}~\bibnamefont
  {Sol{\'e}-Ribalta}}, \bibinfo {author} {\bibfnamefont {A.}~\bibnamefont
  {Arenas}}, \ and\ \bibinfo {author} {\bibfnamefont {S.}~\bibnamefont
  {G{\'o}mez}},\ }\href@noop {} {\bibfield  {journal} {\bibinfo  {journal} {New
  Journal of Physics}\ }\textbf {\bibinfo {volume} {21}},\ \bibinfo {pages}
  {035003} (\bibinfo {year} {2019})}\BibitemShut {NoStop}%
\bibitem [{\citenamefont {Xu}\ \emph {et~al.}(2020)\citenamefont {Xu},
  \citenamefont {Olmos}, \citenamefont {Abbar},\ and\ \citenamefont
  {Gonz{\'a}lez}}]{xu2020deconstructing}%
  \BibitemOpen
  \bibfield  {author} {\bibinfo {author} {\bibfnamefont {Y.}~\bibnamefont
  {Xu}}, \bibinfo {author} {\bibfnamefont {L.~E.}\ \bibnamefont {Olmos}},
  \bibinfo {author} {\bibfnamefont {S.}~\bibnamefont {Abbar}}, \ and\ \bibinfo
  {author} {\bibfnamefont {M.~C.}\ \bibnamefont {Gonz{\'a}lez}},\ }\href@noop
  {} {\bibfield  {journal} {\bibinfo  {journal} {Science Advances}\ }\textbf
  {\bibinfo {volume} {6}},\ \bibinfo {pages} {eabb4112} (\bibinfo {year}
  {2020})}\BibitemShut {NoStop}%
  \bibitem [{\citenamefont {{OpenStreetMap contributors}}(2017)}]{OpenStreetMap}%
  \BibitemOpen
  \bibfield  {author} {\bibinfo {author} {\bibnamefont {{OpenStreetMap
  contributors}}},\ }\href@noop {} {\enquote {\bibinfo {title} {{Planet dump
  retrieved from https://planet.osm.org }},}\ }\bibinfo {howpublished} {\url{
  https://www.openstreetmap.org }} (\bibinfo {year} {2017})\BibitemShut
  {NoStop}%
\bibitem [{\citenamefont {Boeing}(2017)}]{boeing2017osmnx}%
  \BibitemOpen
  \bibfield  {author} {\bibinfo {author} {\bibfnamefont {G.}~\bibnamefont
  {Boeing}},\ }\href@noop {} {\bibfield  {journal} {\bibinfo  {journal}
  {Computers, Environment and Urban Systems}\ }\textbf {\bibinfo {volume}
  {65}},\ \bibinfo {pages} {126} (\bibinfo {year} {2017})}\BibitemShut
  {NoStop}%
\bibitem [{\citenamefont {Orozco}\ \emph {et~al.}(2019)\citenamefont {Orozco},
  \citenamefont {Deritei}, \citenamefont {Vancs{\'o}},\ and\ \citenamefont
  {Vasarhelyi}}]{orozco2019quantifying}%
  \BibitemOpen
  \bibfield  {author} {\bibinfo {author} {\bibfnamefont {L.~G.~N.}\
  \bibnamefont {Orozco}}, \bibinfo {author} {\bibfnamefont {D.}~\bibnamefont
  {Deritei}}, \bibinfo {author} {\bibfnamefont {A.}~\bibnamefont {Vancs{\'o}}},
  \ and\ \bibinfo {author} {\bibfnamefont {O.}~\bibnamefont {Vasarhelyi}},\
  }in\ \href@noop {} {\emph {\bibinfo {booktitle} {International Conference on
  Complex Networks and Their Applications}}}\ (\bibinfo {organization}
  {Springer},\ \bibinfo {year} {2019})\ pp.\ \bibinfo {pages}
  {905--918}\BibitemShut {NoStop}%
\bibitem [{\citenamefont {Boeing}(2019)}]{boeing2019morphology}%
  \BibitemOpen
  \bibfield  {author} {\bibinfo {author} {\bibfnamefont {G.}~\bibnamefont
  {Boeing}},\ }in\ \href@noop {} {\emph {\bibinfo {booktitle} {The mathematics
  of urban morphology}}}\ (\bibinfo  {publisher} {Springer},\ \bibinfo {year}
  {2019})\ pp.\ \bibinfo {pages} {271--287}\BibitemShut {NoStop}%
\bibitem [{\citenamefont {Zhang}\ and\ \citenamefont
  {Zhang}(2019)}]{zhang2019pedestrian}%
  \BibitemOpen
  \bibfield  {author} {\bibinfo {author} {\bibfnamefont {H.}~\bibnamefont
  {Zhang}}\ and\ \bibinfo {author} {\bibfnamefont {Y.}~\bibnamefont {Zhang}},\
  }\href@noop {} {\bibfield  {journal} {\bibinfo  {journal} {Transportation
  research record}\ }\textbf {\bibinfo {volume} {2673}},\ \bibinfo {pages}
  {294} (\bibinfo {year} {2019})}\BibitemShut {NoStop}%
\bibitem [{\citenamefont {Kasemsuppakorn}\ and\ \citenamefont
  {Karimi}(2013)}]{kasemsuppakorn2013pedestrian}%
  \BibitemOpen
  \bibfield  {author} {\bibinfo {author} {\bibfnamefont {P.}~\bibnamefont
  {Kasemsuppakorn}}\ and\ \bibinfo {author} {\bibfnamefont {H.~A.}\
  \bibnamefont {Karimi}},\ }\href@noop {} {\bibfield  {journal} {\bibinfo
  {journal} {Transportation Research part C: emerging technologies}\ }\textbf
  {\bibinfo {volume} {26}},\ \bibinfo {pages} {285} (\bibinfo {year}
  {2013})}\BibitemShut {NoStop}%
\bibitem [{\citenamefont {Mobasheri}\ \emph {et~al.}(2018)\citenamefont
  {Mobasheri}, \citenamefont {Huang}, \citenamefont {Degrossi},\ and\
  \citenamefont {Zipf}}]{mobasheri2018enrichment}%
  \BibitemOpen
  \bibfield  {author} {\bibinfo {author} {\bibfnamefont {A.}~\bibnamefont
  {Mobasheri}}, \bibinfo {author} {\bibfnamefont {H.}~\bibnamefont {Huang}},
  \bibinfo {author} {\bibfnamefont {L.~C.}\ \bibnamefont {Degrossi}}, \ and\
  \bibinfo {author} {\bibfnamefont {A.}~\bibnamefont {Zipf}},\ }\href@noop {}
  {\bibfield  {journal} {\bibinfo  {journal} {Sensors}\ }\textbf {\bibinfo
  {volume} {18}},\ \bibinfo {pages} {509} (\bibinfo {year} {2018})}\BibitemShut
  {NoStop}%
\bibitem [{\citenamefont {Yang}\ \emph {et~al.}(2020)\citenamefont {Yang},
  \citenamefont {Tang}, \citenamefont {Ren}, \citenamefont {Chen},
  \citenamefont {Xie},\ and\ \citenamefont {Li}}]{yang2020pedestrian}%
  \BibitemOpen
  \bibfield  {author} {\bibinfo {author} {\bibfnamefont {X.}~\bibnamefont
  {Yang}}, \bibinfo {author} {\bibfnamefont {L.}~\bibnamefont {Tang}}, \bibinfo
  {author} {\bibfnamefont {C.}~\bibnamefont {Ren}}, \bibinfo {author}
  {\bibfnamefont {Y.}~\bibnamefont {Chen}}, \bibinfo {author} {\bibfnamefont
  {Z.}~\bibnamefont {Xie}}, \ and\ \bibinfo {author} {\bibfnamefont
  {Q.}~\bibnamefont {Li}},\ }\href@noop {} {\bibfield  {journal} {\bibinfo
  {journal} {International Journal of Geographical Information Science}\
  }\textbf {\bibinfo {volume} {34}},\ \bibinfo {pages} {1051} (\bibinfo {year}
  {2020})}\BibitemShut {NoStop}%
\bibitem [{\citenamefont {Ballester}\ \emph {et~al.}(2011)\citenamefont
  {Ballester}, \citenamefont {P{\'e}rez},\ and\ \citenamefont
  {Stuiver}}]{ballester2011automatic}%
  \BibitemOpen
  \bibfield  {author} {\bibinfo {author} {\bibfnamefont {M.~G.}\ \bibnamefont
  {Ballester}}, \bibinfo {author} {\bibfnamefont {M.~R.}\ \bibnamefont
  {P{\'e}rez}}, \ and\ \bibinfo {author} {\bibfnamefont {H.}~\bibnamefont
  {Stuiver}},\ }in\ \href@noop {} {\emph {\bibinfo {booktitle} {Proceedings of
  the 14th AGILE International Conference on Geographic Information Science}}}\
  (\bibinfo {address} {Utrecht, The Netherlands},\ \bibinfo {year}
  {2011})\BibitemShut {NoStop}%
\bibitem [{\citenamefont
  {Kasemsuppakorn}(2011)}]{kasemsuppakorn2011methodology}%
  \BibitemOpen
  \bibfield  {author} {\bibinfo {author} {\bibfnamefont {P.}~\bibnamefont
  {Kasemsuppakorn}},\ }\emph {\bibinfo {title} {Methodology and algorithms for
  pedestrian network construction}},\ \href@noop {} {Ph.D. thesis},\ \bibinfo
  {school} {University of Pittsburgh} (\bibinfo {year} {2011})\BibitemShut
  {NoStop}%
\bibitem [{\citenamefont {Bolten}\ \emph {et~al.}(2015)\citenamefont {Bolten},
  \citenamefont {Amini}, \citenamefont {Hao}, \citenamefont {Ravichandran},
  \citenamefont {Stephens},\ and\ \citenamefont {Caspi}}]{bolten2015urban}%
  \BibitemOpen
  \bibfield  {author} {\bibinfo {author} {\bibfnamefont {N.}~\bibnamefont
  {Bolten}}, \bibinfo {author} {\bibfnamefont {A.}~\bibnamefont {Amini}},
  \bibinfo {author} {\bibfnamefont {Y.}~\bibnamefont {Hao}}, \bibinfo {author}
  {\bibfnamefont {V.}~\bibnamefont {Ravichandran}}, \bibinfo {author}
  {\bibfnamefont {A.}~\bibnamefont {Stephens}}, \ and\ \bibinfo {author}
  {\bibfnamefont {A.}~\bibnamefont {Caspi}},\ }in\ \href@noop {} {\emph
  {\bibinfo {booktitle} {Proceedings of the 1st International ACM SIGSPATIAL
  Workshop on Smart Cities and Urban Analytics}}}\ (\bibinfo {year} {2015})\
  pp.\ \bibinfo {pages} {122--125}\BibitemShut {NoStop}%
\bibitem [{\citenamefont {Csardi}\ and\ \citenamefont
  {Nepusz}(2006)}]{Csardi:2006aa}%
  \BibitemOpen
  \bibfield  {author} {\bibinfo {author} {\bibfnamefont {G.}~\bibnamefont
  {Csardi}}\ and\ \bibinfo {author} {\bibfnamefont {T.}~\bibnamefont
  {Nepusz}},\ }\href {http://igraph.org} {\bibfield  {journal} {\bibinfo
  {journal} {InterJournal}\ }\textbf {\bibinfo {volume} {Complex Systems}},\
  \bibinfo {pages} {1695} (\bibinfo {year} {2006})}\BibitemShut {NoStop}%
\end{thebibliography}%


\begin{thebibliography}{12}%
\makeatletter
\providecommand \@ifxundefined [1]{%
 \@ifx{#1\undefined}
}%
\providecommand \@ifnum [1]{%
 \ifnum #1\expandafter \@firstoftwo
 \else \expandafter \@secondoftwo
 \fi
}%
\providecommand \@ifx [1]{%
 \ifx #1\expandafter \@firstoftwo
 \else \expandafter \@secondoftwo
 \fi
}%
\providecommand \natexlab [1]{#1}%
\providecommand \enquote  [1]{``#1''}%
\providecommand \bibnamefont  [1]{#1}%
\providecommand \bibfnamefont [1]{#1}%
\providecommand \citenamefont [1]{#1}%
\providecommand \href@noop [0]{\@secondoftwo}%
\providecommand \href [0]{\begingroup \@sanitize@url \@href}%
\providecommand \@href[1]{\@@startlink{#1}\@@href}%
\providecommand \@@href[1]{\endgroup#1\@@endlink}%
\providecommand \@sanitize@url [0]{\catcode `\\12\catcode `\$12\catcode
  `\&12\catcode `\#12\catcode `\^12\catcode `\_12\catcode `\%12\relax}%
\providecommand \@@startlink[1]{}%
\providecommand \@@endlink[0]{}%
\providecommand \url  [0]{\begingroup\@sanitize@url \@url }%
\providecommand \@url [1]{\endgroup\@href {#1}{\urlprefix }}%
\providecommand \urlprefix  [0]{URL }%
\providecommand \Eprint [0]{\href }%
\providecommand \doibase [0]{http://dx.doi.org/}%
\providecommand \selectlanguage [0]{\@gobble}%
\providecommand \bibinfo  [0]{\@secondoftwo}%
\providecommand \bibfield  [0]{\@secondoftwo}%
\providecommand \translation [1]{[#1]}%
\providecommand \BibitemOpen [0]{}%
\providecommand \bibitemStop [0]{}%
\providecommand \bibitemNoStop [0]{.\EOS\space}%
\providecommand \EOS [0]{\spacefactor3000\relax}%
\providecommand \BibitemShut  [1]{\csname bibitem#1\endcsname}%
\let\auto@bib@innerbib\@empty
\bibitem [{\citenamefont {{OpenStreetMap contributors}}(2017)}]{OpenStreetMap}%
  \BibitemOpen
  \bibfield  {author} {\bibinfo {author} {\bibnamefont {{OpenStreetMap
  contributors}}},\ }\href@noop {} {\enquote {\bibinfo {title} {{Planet dump
  retrieved from https://planet.osm.org }},}\ }\bibinfo {howpublished} {\url{
  https://www.openstreetmap.org }} (\bibinfo {year} {2017})\BibitemShut
  {NoStop}%
\bibitem [{\citenamefont {Boeing}(2017)}]{boeing2017osmnx}%
  \BibitemOpen
  \bibfield  {author} {\bibinfo {author} {\bibfnamefont {G.}~\bibnamefont
  {Boeing}},\ }\href@noop {} {\bibfield  {journal} {\bibinfo  {journal}
  {Computers, Environment and Urban Systems}\ }\textbf {\bibinfo {volume}
  {65}},\ \bibinfo {pages} {126} (\bibinfo {year} {2017})}\BibitemShut
  {NoStop}%
\bibitem [{\citenamefont {Orozco}\ \emph {et~al.}(2019)\citenamefont {Orozco},
  \citenamefont {Deritei}, \citenamefont {Vancs{\'o}},\ and\ \citenamefont
  {Vasarhelyi}}]{orozco2019quantifying}%
  \BibitemOpen
  \bibfield  {author} {\bibinfo {author} {\bibfnamefont {L.~G.~N.}\
  \bibnamefont {Orozco}}, \bibinfo {author} {\bibfnamefont {D.}~\bibnamefont
  {Deritei}}, \bibinfo {author} {\bibfnamefont {A.}~\bibnamefont {Vancs{\'o}}},
  \ and\ \bibinfo {author} {\bibfnamefont {O.}~\bibnamefont {Vasarhelyi}},\
  }in\ \href@noop {} {\emph {\bibinfo {booktitle} {International Conference on
  Complex Networks and Their Applications}}}\ (\bibinfo {organization}
  {Springer},\ \bibinfo {year} {2019})\ pp.\ \bibinfo {pages}
  {905--918}\BibitemShut {NoStop}%
\bibitem [{\citenamefont {Boeing}(2019)}]{boeing2019morphology}%
  \BibitemOpen
  \bibfield  {author} {\bibinfo {author} {\bibfnamefont {G.}~\bibnamefont
  {Boeing}},\ }in\ \href@noop {} {\emph {\bibinfo {booktitle} {The mathematics
  of urban morphology}}}\ (\bibinfo  {publisher} {Springer},\ \bibinfo {year}
  {2019})\ pp.\ \bibinfo {pages} {271--287}\BibitemShut {NoStop}%
\bibitem [{\citenamefont {Zhang}\ and\ \citenamefont
  {Zhang}(2019)}]{zhang2019pedestrian}%
  \BibitemOpen
  \bibfield  {author} {\bibinfo {author} {\bibfnamefont {H.}~\bibnamefont
  {Zhang}}\ and\ \bibinfo {author} {\bibfnamefont {Y.}~\bibnamefont {Zhang}},\
  }\href@noop {} {\bibfield  {journal} {\bibinfo  {journal} {Transportation
  research record}\ }\textbf {\bibinfo {volume} {2673}},\ \bibinfo {pages}
  {294} (\bibinfo {year} {2019})}\BibitemShut {NoStop}%
\bibitem [{\citenamefont {Kasemsuppakorn}\ and\ \citenamefont
  {Karimi}(2013)}]{kasemsuppakorn2013pedestrian}%
  \BibitemOpen
  \bibfield  {author} {\bibinfo {author} {\bibfnamefont {P.}~\bibnamefont
  {Kasemsuppakorn}}\ and\ \bibinfo {author} {\bibfnamefont {H.~A.}\
  \bibnamefont {Karimi}},\ }\href@noop {} {\bibfield  {journal} {\bibinfo
  {journal} {Transportation Research part C: emerging technologies}\ }\textbf
  {\bibinfo {volume} {26}},\ \bibinfo {pages} {285} (\bibinfo {year}
  {2013})}\BibitemShut {NoStop}%
\bibitem [{\citenamefont {Mobasheri}\ \emph {et~al.}(2018)\citenamefont
  {Mobasheri}, \citenamefont {Huang}, \citenamefont {Degrossi},\ and\
  \citenamefont {Zipf}}]{mobasheri2018enrichment}%
  \BibitemOpen
  \bibfield  {author} {\bibinfo {author} {\bibfnamefont {A.}~\bibnamefont
  {Mobasheri}}, \bibinfo {author} {\bibfnamefont {H.}~\bibnamefont {Huang}},
  \bibinfo {author} {\bibfnamefont {L.~C.}\ \bibnamefont {Degrossi}}, \ and\
  \bibinfo {author} {\bibfnamefont {A.}~\bibnamefont {Zipf}},\ }\href@noop {}
  {\bibfield  {journal} {\bibinfo  {journal} {Sensors}\ }\textbf {\bibinfo
  {volume} {18}},\ \bibinfo {pages} {509} (\bibinfo {year} {2018})}\BibitemShut
  {NoStop}%
\bibitem [{\citenamefont {Yang}\ \emph {et~al.}(2020)\citenamefont {Yang},
  \citenamefont {Tang}, \citenamefont {Ren}, \citenamefont {Chen},
  \citenamefont {Xie},\ and\ \citenamefont {Li}}]{yang2020pedestrian}%
  \BibitemOpen
  \bibfield  {author} {\bibinfo {author} {\bibfnamefont {X.}~\bibnamefont
  {Yang}}, \bibinfo {author} {\bibfnamefont {L.}~\bibnamefont {Tang}}, \bibinfo
  {author} {\bibfnamefont {C.}~\bibnamefont {Ren}}, \bibinfo {author}
  {\bibfnamefont {Y.}~\bibnamefont {Chen}}, \bibinfo {author} {\bibfnamefont
  {Z.}~\bibnamefont {Xie}}, \ and\ \bibinfo {author} {\bibfnamefont
  {Q.}~\bibnamefont {Li}},\ }\href@noop {} {\bibfield  {journal} {\bibinfo
  {journal} {International Journal of Geographical Information Science}\
  }\textbf {\bibinfo {volume} {34}},\ \bibinfo {pages} {1051} (\bibinfo {year}
  {2020})}\BibitemShut {NoStop}%
\bibitem [{\citenamefont {Ballester}\ \emph {et~al.}(2011)\citenamefont
  {Ballester}, \citenamefont {P{\'e}rez},\ and\ \citenamefont
  {Stuiver}}]{ballester2011automatic}%
  \BibitemOpen
  \bibfield  {author} {\bibinfo {author} {\bibfnamefont {M.~G.}\ \bibnamefont
  {Ballester}}, \bibinfo {author} {\bibfnamefont {M.~R.}\ \bibnamefont
  {P{\'e}rez}}, \ and\ \bibinfo {author} {\bibfnamefont {H.}~\bibnamefont
  {Stuiver}},\ }in\ \href@noop {} {\emph {\bibinfo {booktitle} {Proceedings of
  the 14th AGILE International Conference on Geographic Information Science}}}\
  (\bibinfo {address} {Utrecht, The Netherlands},\ \bibinfo {year}
  {2011})\BibitemShut {NoStop}%
\bibitem [{\citenamefont
  {Kasemsuppakorn}(2011)}]{kasemsuppakorn2011methodology}%
  \BibitemOpen
  \bibfield  {author} {\bibinfo {author} {\bibfnamefont {P.}~\bibnamefont
  {Kasemsuppakorn}},\ }\emph {\bibinfo {title} {Methodology and algorithms for
  pedestrian network construction}},\ \href@noop {} {Ph.D. thesis},\ \bibinfo
  {school} {University of Pittsburgh} (\bibinfo {year} {2011})\BibitemShut
  {NoStop}%
\bibitem [{\citenamefont {Bolten}\ \emph {et~al.}(2015)\citenamefont {Bolten},
  \citenamefont {Amini}, \citenamefont {Hao}, \citenamefont {Ravichandran},
  \citenamefont {Stephens},\ and\ \citenamefont {Caspi}}]{bolten2015urban}%
  \BibitemOpen
  \bibfield  {author} {\bibinfo {author} {\bibfnamefont {N.}~\bibnamefont
  {Bolten}}, \bibinfo {author} {\bibfnamefont {A.}~\bibnamefont {Amini}},
  \bibinfo {author} {\bibfnamefont {Y.}~\bibnamefont {Hao}}, \bibinfo {author}
  {\bibfnamefont {V.}~\bibnamefont {Ravichandran}}, \bibinfo {author}
  {\bibfnamefont {A.}~\bibnamefont {Stephens}}, \ and\ \bibinfo {author}
  {\bibfnamefont {A.}~\bibnamefont {Caspi}},\ }in\ \href@noop {} {\emph
  {\bibinfo {booktitle} {Proceedings of the 1st International ACM SIGSPATIAL
  Workshop on Smart Cities and Urban Analytics}}}\ (\bibinfo {year} {2015})\
  pp.\ \bibinfo {pages} {122--125}\BibitemShut {NoStop}%
\bibitem [{\citenamefont {Csardi}\ and\ \citenamefont
  {Nepusz}(2006)}]{Csardi:2006aa}%
  \BibitemOpen
  \bibfield  {author} {\bibinfo {author} {\bibfnamefont {G.}~\bibnamefont
  {Csardi}}\ and\ \bibinfo {author} {\bibfnamefont {T.}~\bibnamefont
  {Nepusz}},\ }\href {http://igraph.org} {\bibfield  {journal} {\bibinfo
  {journal} {InterJournal}\ }\textbf {\bibinfo {volume} {Complex Systems}},\
  \bibinfo {pages} {1695} (\bibinfo {year} {2006})}\BibitemShut {NoStop}%
\end{thebibliography}%
%

\end{document}


\title{Supplementary Material\\ Planning for sustainable Open Streets in pandemic cities}

\author{Daniel Rhoads}
\affiliation{Internet Interdisciplinary Institute (IN3), Universitat Oberta de Catalunya, Barcelona, Spain}

\author{Albert Sol\'e-Ribalta}
\affiliation{Internet Interdisciplinary Institute (IN3), Universitat Oberta de Catalunya, Barcelona, Catalonia, Spain}
\affiliation{URPP Social Networks, Universit\"at Z\"urich, Z\"urich, Switzerland}

\author{Marta C. Gonz\'alez}
\affiliation{Department of City and Regional Planning, University of California, Berkeley, CA 94720, USA}
\affiliation{Energy Technologies Area, Lawrence Berkeley National Laboratory, Berkeley, CA 94720, USA}
\affiliation{Department of Civil and Environmental Engineering, University of California, Berkeley, CA 94720, USA}

\author{Javier Borge-Holthoefer}
\affiliation{Internet Interdisciplinary Institute (IN3), Universitat Oberta de Catalunya, Barcelona, Catalonia, Spain}

\date{\today}

\maketitle


\section{Materials and Methods}

\subsection{Materials}

\subsubsection{Open data on public space}

As mentioned in the main text, public data on sidewalk infrastructure is not standardised to the same extent as road network data. Many municipal open data portals lack clearly available sidewalk data entirely. When available, sidewalk datasets generally take the form either of sets of  sidewalk centerlines, or sets of sidewalk polygons. The latter is considerably more common, and also provides implicit information on the width of sidewalks. For this reason, the sidewalk data gathered for this work was restricted to cities with an available sidewalk polygon dataset. However, the algorithm for network construction could be easily adapted for other types of sidewalk data.

Table~\ref{table:sources} lists the public sources from which sidewalk geometries have been collected. Note that the formats in which these data are encoded varies from city to city.

\begin{table*}[th]
  \centering
\begin{tabular}{c|c}
\hline
\bf{City} & \bf{Data source} \\
\hline
\hline
Denver & https://www.denvergov.org/opendata \\
Montreal & http://donnees.ville.montreal.qc.ca \\
Washington D.C. & https://opendata.dc.gov \\
Boston & https://data.boston.gov \\
New York & https://opendata.cityofnewyork.us \\
Buenos Aires &https://data.buenosaires.gob.ar \\
Bogotá & https://datosabiertos.bogota.gov.co \\
Brussels & https://datastore.brussels/web/data \\
Barcelona & http://www.icc.cat/vissir3 \\
Paris & https://opendata.paris.fr/pages/home \\ 
\end{tabular}
  \caption{List of open data portals used to acquire public space data for the 10 cities of study.} 
  \label{table:sources}
\end{table*}

\subsubsection{Road networks}
Road network geometries were extracted from OpenStreetMap (OSM) \cite{OpenStreetMap} using the OSMnx Python package \cite{boeing2017osmnx}, which provides a simple interface for querying OSM data. The package was used to extract the edges and nodes of the ``Drive'' and ``Walk'' networks of each city. The ``Drive'' network was used as a basis for sidewalk network construction, and later as the road network for the purposes of the Street Pedestrianization process (see final two sections below). The ``Walk'' network, which was filtered to only include those edges with a ``highway'' tag of ``pedestrian'', ``path'', or ``living street'', was incorporated into the Sidewalk Network.

\subsection{Methods}

\subsubsection{Street space distribution calculations}
For the purposes of this study, urban space was divided into 4 broad categories: Road, Sidewalk, Buildings, and Parkland. The focus of our analysis of public space was on the distribution of street space between cars and pedestrians. Accordingly, Buildings and Parkland were neglected in this work.

For the purposes of calculating the area taken up by Road, datasets of roadbed polygons were used, when available. Otherwise, roadbed polygons were calculated as the area left over after subtracting all other categories of public space from the city (Buildings, Parkland, and Sidewalk). 

\subsubsection{Sidewalk networks}

\paragraph{Literature review on sidewalk network construction}
Several attempts have been made in the past to model sidewalk infrastructure in network form \cite{orozco2019quantifying,boeing2019morphology,zhang2019pedestrian,kasemsuppakorn2013pedestrian,mobasheri2018enrichment,yang2020pedestrian,ballester2011automatic,kasemsuppakorn2011methodology,bolten2015urban}.

Considering the adjacency of sidewalks to roads, the simplest is to use the road network itself \cite{orozco2019quantifying,boeing2019morphology}. In this case, nodes and edges generally represent intersections and street segments respectively, as is usual in a road network. This most basic representation assumes the presence of sidewalks along the edges of every street segment, which is a serious exaggeration. To make this approach more realistic, information about sidewalk presence can be encoded as link meta-data. Sometimes, this means a boolean value indicating whether or not that edge is traversable by at least once sidewalk. Some more detailed approaches (the OpenStreetMap working standard, for example) occasionally indicate the presence of sidewalks on both, one, or neither sides. However, this abstracts away from geometric differences between sidewalks and road segments, as well as the costs associated with sidewalk crossings. The more information that is added to the street network representation, the more plain it becomes that a distinct object is more apt to the task of modelling sidewalk networks.

Several works have treated sidewalk networks in this way, as a dedicated structure separate from the road network. Some such studies have been based on hand-drawn networks based on field data or aerial imagery \cite{zhang2019pedestrian}. This type of data capture tends to be quite reliable, but is labor intensive and difficult to scale. Another approach is to crowd-source sidewalk data, making use of pedestrian GPS traces to identify real walking paths \cite{kasemsuppakorn2013pedestrian,mobasheri2018enrichment,yang2020pedestrian}. This has a number of pitfalls, such as lack of data availability and exposure to error, and may be more interesting for the study of pedestrian dynamics than sidewalk infrastructure. The road network can also be geometrically buffered to generate more realistic sidewalk geometries without empirical sidewalk data \cite{ballester2011automatic,kasemsuppakorn2011methodology}, but this method can overestimate connectivity.

Finally, there have been a few studies that have attempted to generate sidewalk networks in an automatic fashion from municipal sidewalk data \cite{bolten2015urban}. In general, these have been limited both geographically and by the relative diversity of formats of municipal sidewalk data. At present, while a standardized road network can be downloaded with a few clicks or lines of code, in both network and GIS formats, there is no widely-accepted format for sidewalk data. Broadly speaking, sidewalk data is a seriously neglected subset of urban data, and is often lost and forgotten between ubiquitous street networks and detailed cadastral maps. Where sidewalk data is available, its most common form is as a feature in planimetric datasets, where sidewalks are represented as polygons. This is a very literal representation, but it has some advantages as a starting point for constructing pedestrian networks, particularly since it implies information on sidewalk width.

\paragraph{Network construction algorithm}
The network construction algorithm requires two sources of inputs: sidewalk geometries and road network geometries. All geometries were saved in ESRI ShapeFile format.

Sidewalk geometries were acquired as described above, in polygon form. The exterior line of each polygon was extracted, to serve as the linear basis for the links of the network. These line geometries were cleaned to ensure contiguity with the road edge. Each sidewalk geometry was then assigned to a city block, defined as a graph face formed by a closed loop of segments from the road network (for more information on the road networks used, see the road network section above). As stated in the main text, nodes of the sidewalk network are placed at every point where a pedestrian might move from one block to another, and at points of change in direction. Thus, for each city block, one node is placed for each intersection associated with that block. This node is placed at the point along the sidewalk geometry closest to the intersection in question. Additional nodes are placed along the sidewalk geometries at vertices where the sidewalk changes direction. 

The nodes are then used to ``cut'' the sidewalk geometries into shorter segments, representing the links of the network, through an iterative process. The link geometries are labelled according to the nodes that they touch. Finally, nodes from adjacent blocks are connected according to several rules: first, that the crosswalk links only cross one segment of the road Network, and second, that they do not touch other crosswalk geometries except at network nodes.

As noted in the road network section, data on pedestrian streets from OpenStreetMap is also downloaded using the OSMnx Python package. The geometries are represented as street center lines, as in the classic road Network representation. These geometries are overlaid onto the geometries of the Sidewalk Network and incorporated as links and nodes. 

Some basic statistics on the networks constructed for Paris and New York are shown in Table~\ref{table:networks}.

\begin{table*}[th]
  \centering
\begin{tabular}{c|c|c|c}
\hline
\bf{City} & \bf{Metric} & {\bf Sidewalk network} & {\bf Road network} \\
\hline
\hline
\multirow{ 6}{*}{\it{Paris}} & $N$ & 39536 & 10073 \\
& $L$ & 67374 & 19651 \\
& $\langle k \rangle$ & 3.41 &  1.95\\
& $C$ & 0.11 & 0.07 \\
& $\langle E \rangle$ & 1.1e-3 & 2.2e-3 \\
& $\delta$ & 23300 meters & 19533 meters \\
\hline
\multirow{ 6}{*}{\it{New York}}  &$N$ & 155811 & 55318 \\
& $L$ & 284524 & 140468 \\
& $\langle k \rangle$ & 3.65 & 2.54 \\
& $C$ & 0.08 & 0.04 \\
& $\langle E \rangle$ & 3.2e-5 & 7.9e-5 \\
& $\delta$ & 62148 meters & 67269 meters \\
\hline
\end{tabular}
  \caption{Comparison of pedestrian and road networks. $N$ and $L$ denote the number of nodes and links of the graph respectively. $\langle k \rangle$ is the mean degree. $C$ is the clustering coefficient of the graph. $\langle E \rangle$ is its average efficiency. $\delta$ is the graph's diameter.} 
  \label{table:networks}
\end{table*}

\paragraph{Sidewalk width calculation}
Sidewalk widths were calculated using a simple heuristic based on the geometries of the sidewalk network links and the associated sidewalk polygons from which the links were derived. Each link of a particular block was assigned its ``share'' of the associated sidewalk polygon using Voronoi polygons. The area of this share was then divided by the length of the link to acquire an average width for that link. An alternative would have been to identify the minimum width of each segment, but this would be more sensitive to errors in the original sidewalk data, and could lead to an under-estimation of the network's resilience to the width percolation process.

\paragraph{Width percolation}
As described in the paper, for the purposes of percolation, the links of the sidewalk network were sorted by edge width from lowest to highest (excluding pedestrian streets and crosswalks, whose widths are set to -1). Accordingly, the first edge to be percolated is the narrowest sidewalk in the network, and percolated edges become progressively wider from there. Networks were imported from geometric (ShapeFile) format into the Python implementation of the Igraph library \cite{Csardi:2006aa}. 

\subsubsection{Connecting the sidewalk and road networks}
In order to carry out the ``Open Streets'' pacification process described in the results, and in the sections below, an interface between the road and Sidewalk Networks needed to be made. Given that the Sidewalk Network is constructed using the geometries of the Road Network, this is relatively straightforward. Each edge of the Sidewalk Network is adjacent to a corresponding edge of the Road Network. Similarly to the calculation of sidewalk widths, Voronoi polygons based on the road edges were used to assign each Sidewalk link to its street segment. In this way, each edge of the Sidewalk Network has one corresponding edge from the Road Network, but one edge of the Road Network may be assigned to multiple Sidewalk edges. 

\subsubsection{``Open Streets'' pacification process}
\label{betweenness}
To implement the street pacification process, an extra step was added to the original width percolation algorithm. As described in section S2.d. above, each edge of the sidewalk network is associated with one edge of the road network. A test is performed in which both the sidewalk edge in question, and its corresponding road network edge, are removed from their respective networks. The resulting change in the size of the giant component of both networks (sidewalk and road) is compared. The algorithm then chooses to permanently delete the edge whose removal led to a smaller proportional decrease in giant component size for its network. The other edge is left in place. When an edge of the road network is removed, all of its corresponding sidewalks have their widths set to -1, excluding them from future removal. 

In the case of a tie (i.e., the change in component size is equal for both networks), several different approaches were considered. On the simpler side, one could opt to always favor the sidewalk network, making an essentially normative choice. Another option would be to effectively ``flip a coin'', i.e. decide which link to remove by a Bernoulli trial. 

It was finally decided that, in the case of a tie, there should be an attempt to optimize another variable of interest: the average path length of the road network. Thus, using our algorithm, in the event of a tie (same loss for both interdependent structures), the sidewalk is removed with probability $p \sim B_{ij}$ , where $B_{ij}$ is the time-weighted edge betweenness (see below) of road segment $(i, j)$; conversely, the road segment $(i, j)$ is removed with probability $1 - p$. Since an edge's betweenness is calculated based on the number of shortest paths passing through it, betweenness can be taken as a proxy for an edge's usefulness in keeping path lengths low. This method produced satisfactory results. 

\subsubsection{Time-weighted betweenness on road networks}
Road networks are generally treated as weighted, directed networks in the literature, and our work is no exception. While speed on a road network can be modelled as constant, taking only the length of road links as a weight, in the real world some road segments permit much faster travel than others (i.e. major arteries, highways). 

To incorporate this, our road networks are weighted by time (as opposed to sheer distance). Some OpenStreetMap street segments have a data attribute indicating their posted speed limit. When this was available, it was used. Otherwise, speeds were determined by the segment's ``highway'' attribute, which establishes a general typology of segments (see Table~\ref{table:speeds}). If no `highway'' attribute was given, speed was assumed to be 50 km/h.

\begin{table*}[th]
  \centering
\begin{tabular}{c|c}
\hline
\bf{OSM ``highway'' attribute} & \bf{Speed (Km/H)} \\
\hline
\hline
Motorway &130 \\
Trunk & 130 \\
Primary & 90 \\
Secondary & 80 \\
Tertiary & 80 \\
Unclassified & 60 \\
Road & 60 \\
Residential & 25 \\
motorway\textunderscore link & 90 \\
trunk\textunderscore link & 90 \\ 
primary\textunderscore link & 90 \\
secondary\textunderscore link &80 \\
tertiary\textunderscore link &80 \\
living\textunderscore street & 25 \\
construction & 50 \\
service & 25 \\ 
\end{tabular}
  \caption{{\bf Mapping of OpenStreetMap “highway” tags to speeds in Km/H.} All road network links with a given “highway” attribute were assigned the corresponding speed as a weight, for the purposes of calculating shortest paths and time-weighted betweenness. }
  \label{table:speeds}
\end{table*}

The time necessary to traverse a link was then taken to be simply the length of the link divided by the speed limit. 

\subsubsection{Calculating change in street space distribution, post-intervention}
As described in the text, blocking traffic to convert a road segment into an Open Street necessarily increases the amount of street space dedicated to pedestrians. To calculate the change in overall street space distribution, we first had to assign to each road segment its share of road space. This was done again by means of voronoi polygons. Thus, each road segment was associated with that part of the total road space that was closer to it than to any other segment. When a road segment is converted to an Open Street by our heuristic process, we henceforth consider its assigned road space to be ``sidewalk'' for the purposes of calculating the distribution of street space

\section{Supplementary Text}

\subsection{Width-based percolation and street pacification for other cities}

We provide here the complete analysis, performed for Paris and New York in the main text, corresponding to the additional eight cities whose ``arrogance of space'' is reported in Figure 1. These are: Denver (Figure S1), Montreal (Figure S2), Washington D.C. (Figure S3), Boston (Figure S4), Buenos Aires (Figure S5), Bogotá (Figure S6), Brussels (Figure S7), Barcelona (Figure S8). Each of these supplemental Figures includes:

\begin{itemize}
\item Worst-case scenario: no intervention is proposed, and edge removal is solely based on the width of sidewalks. (corresponds to left panel of Figure 3 in the main text)
\item Map representation of the sidewalk network's connected components, at $\tau$ = 0 and $\tau$ = 5 meters thresholds. (corresponds to right panel of Figure 3 in the main text)
\item Evolution of connected components under the proposed ``Open Streets'' shared-effort heuristic. (corresponds to left panel of Figure 4 in the main text) 
\item Deterioration of shortest paths (travel cost) for the road network, as a consequence of the application of the ``Open Streets'' shared-effort heuristic. (corresponds to right panel of Figure 4 in the main text)
\end{itemize}

Overall, the results in the main text and Figs.S1-S8 are quite heterogeneous. Attending the no-intervention scenario only, we find a majority of cities whose sidewalk network’s current state is extremely fragile: Denver, Montreal, Washington D.C., Boston, Bogotá, Brussels, and New York to some extent. All these cities’ sidewalk networks reach the percolation critical point (SG peak) before a threshold of 3.5 meters, making them unable to confront the challenges of social distancing while preserving a connected pedestrian infrastructure. Only Buenos Aires, Barcelona and Paris present a robust sidewalk network beyond  $\tau$ = 3.5.

Then, under the ``Open Streets'' heuristic, one would expect that fragile cities would render similar (and probably poor) results. Interestingly, this is not the case. Some cities, like New York (in the main text), Washington D.C. (Fig. S3), Bogotá (Fig. S6) and Brussels (Fig. S7), seem to respond very well to the shared-effort strategy, at least up to $\tau$ = 3.5. On the contrary, although delayed to some extent, the collapse of $G^*_s$ for cities like Denver, Montreal or Boston happens early, and a connected sidewalk network that meets social distancing constraints seems difficult to attain. Some insights on these diverse outcomes can be found in the particular spatial distribution of sidewalk widths across a city, and its relationship with the centrality (betweenness) of the corresponding road segments. Next section and Figure S9 are devoted to such relationships.

\subsection{Correlation of sidewalk width to road edge betweenness}

To understand the possible reasons why, in some cities with similar percolation profiles, the shared-effort heuristic renders different improvements to sustain the sidewalk network, one has to review the assumptions of the heuristic itself. Particularly, the heuristic assumes that, whenever a tie happens, it is solved considering the road segment edge betweenness Bij: the larger Bij, the more probable it is to save that road segment (see main text and Section S.2.e above). In practice, this implies that a preference is in place to preserve the most critical parts of the road network (and thus its overall functionality).

Adding to the previous point, we need to keep in mind that, for a social-distancing-compliant sidewalk network, the heuristic needs to preserve many sidewalk segments between 2 and 4 meters: the ones that can delay the collapse of the network to an acceptable level. 

Putting these two ideas together, we can arrive at a possible explanation: the heuristic works very well if the city includes many sidewalk widths between 2 and 4 meters, and if those same sidewalks are adjacent to low betweenness road segments. Whenever both circumstances happen, in the face of a tie, the percolation process will, with high probability, choose to block the segment to cars (and preserve the sidewalk). If that happens infrequently, then the heuristic will tend to favor the road network, causing a rapid collapse in sidewalk network connectivity.

These ideas can be better understood visually. The scatter plots in Figure S9 evince that there is no identifiable pattern that correlates sidewalk width with road centrality. Rather, each city presents its own and unique profile. If anything, it is possible to conclude that, typically, the widest sidewalks can be found in high-betweenness road segments. This lack of correlation is expected, given that there is no a priori necessity (nor obligation) for city governments to plan the width of sidewalks as a function of road centrality. Note that we have informally sorted the scatter plots, from cities in which the heuristic was more (top)  to less (bottom) successful.

Beyond these observations, the important aspect of Fig. S9 is to highlight that the least successful cities (bottom row) lack precisely low betweenness roads at middle ranges of sidewalk width (marked with a red square). On the opposite side (top row), Paris, Barcelona and Buenos Aires present a robust sidewalk network, and this has no relation to the heuristic nor to road centralities, but rather to a rich distribution of wide sidewalks. In the middle, we find New York, Washington D.C., Bogotá and Brussels: these are cities whose no-intervention percolation process rendered a pessimistic outlook (early network collapse), but then respond very well to the shared-effort strategy.  The common trait for these cities is precisely a rich presence of low-betweenness roadbeds in the range 2-4 meters (red square).

\begin{figure}[h!]
\centering
	\includegraphics[width=.80\textwidth]{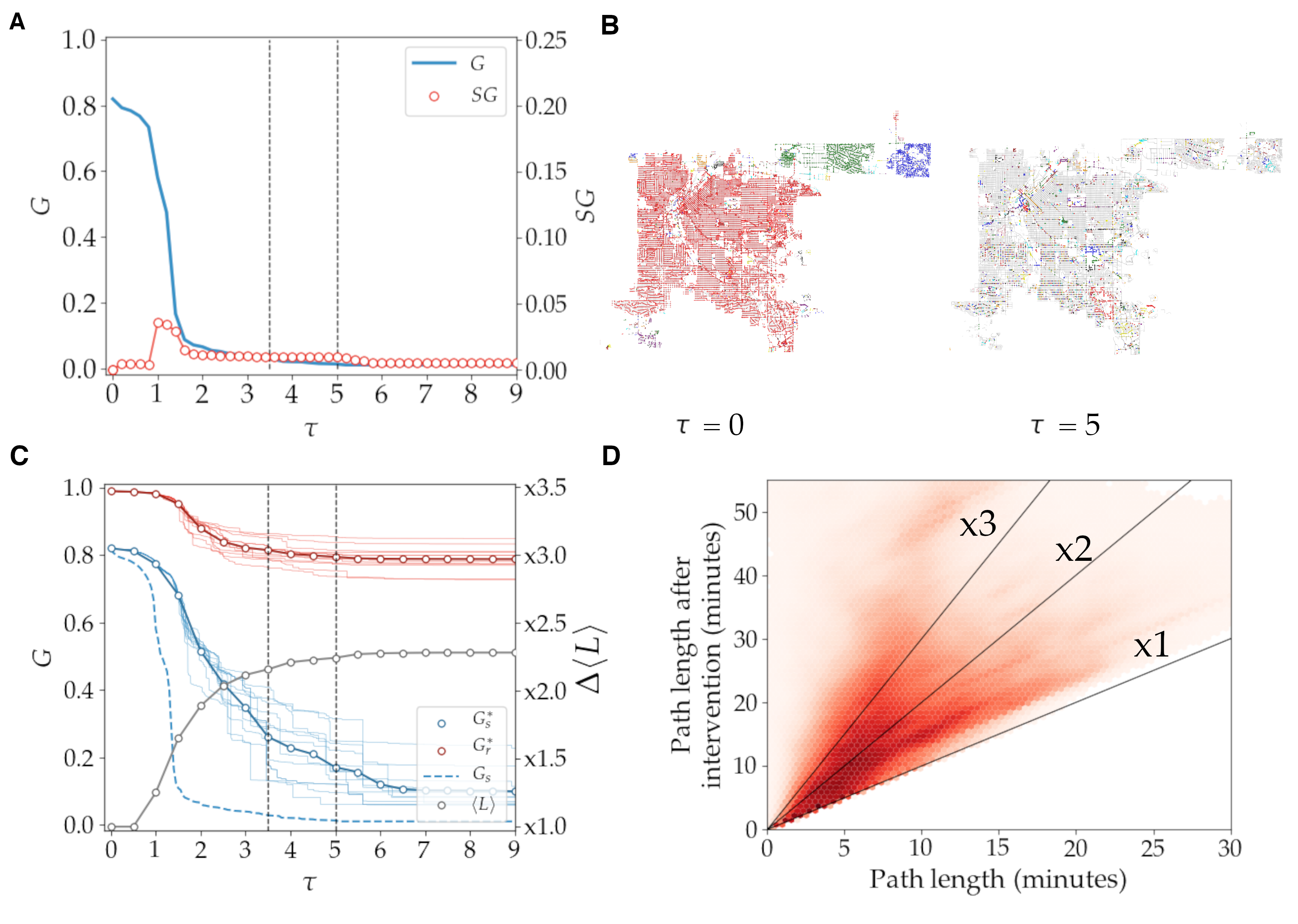}
\caption{{\bf Social distancing in Denver's sidewalk network.}  (A) The evolution of G (solid blue), and the peak of SG (red circles) at $\tau$ = 1 meter indicate that the sidewalk network for this city presents a weak connectivity, and cannot withstand even a moderate requirement of social distance. (B) As for the case of New York City (Fig. 3 of the main text), Denver also presents a fragmented scenario at $\tau$ = 0 (note at least three large connected components); at $\tau$ = 5, there is no single component comparable to the size of the network. (C) The proposed ``Open Streets'' heuristic (solid blue) renders some improvement with respect to the baseline, no-intervention scenario (dashed blue), with $G^*_s$ = 0.25 at $\tau$ = 3.5, and $G^*_s$ $\approx$ 0.2 at $\tau$ = 5. The cost of such improvement is on the road network, whose giant component $G^*_r$ is deteriorated by 20\%. (D) The distribution of the increase in average path lengths (in minutes) for drivers when the process has been run up to $\tau$ = 5 shows that shorter trips (up to 10 minutes) may triple as a consequence of the pedestrianization process. Most longer trips (> 10 minutes), however, see a moderate increase by less than a factor of 2.}
\label{fig:denver}
\end{figure}

\begin{figure}[h!]
\centering
	\includegraphics[width=.80\textwidth]{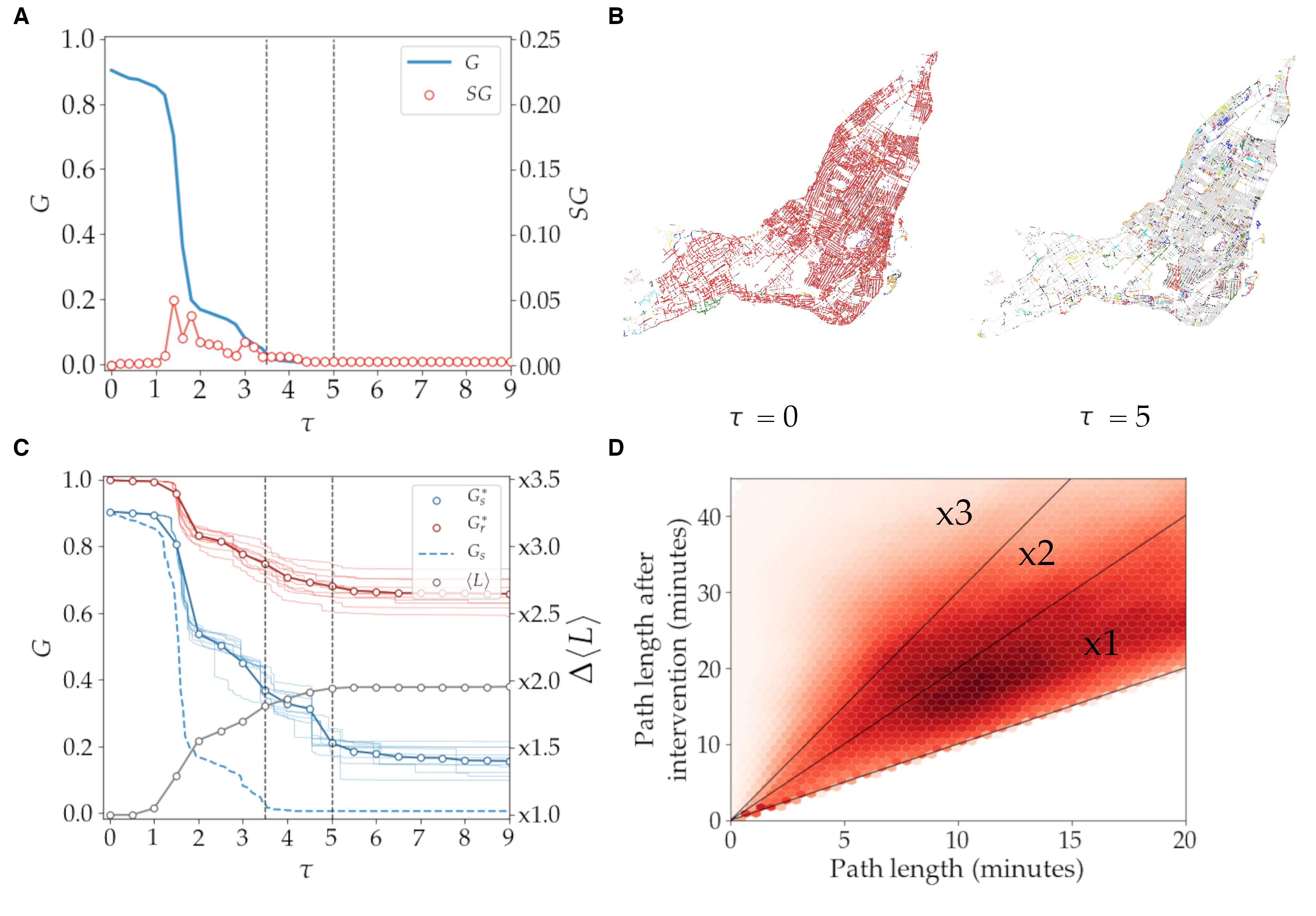}
\caption{{\bf NSocial distancing in Montreal's sidewalk network.} (A) The evolution of G (solid blue), and the peak of SG (red circles) at $\tau$ = 1.5 meters indicate that the sidewalk network for this city presents a weak connectivity, and cannot withstand even a moderate requirement of social distance. (B) As for the case of New York City (Fig. 3 of the main text), Montreal also presents a fragmented scenario at $\tau$ = 0, but the disconnected subnetworks are scattered in the outer periphery of the city. At $\tau$ = 5 there is no single component comparable to the size of the network. (C) The proposed ``Open Streets'' heuristic (solid blue) renders a notable improvement with respect to the baseline, no-intervention scenario (dashed blue), with $G^*_s$ $\approx$ 0.4 at $\tau$ = 3.5, and $G^*_s$ = 0.2 at $\tau$ = 5. The cost of such improvement is on the road network, whose giant component $G^*_r$ is deteriorated by 20 to 25\% (at 3.5 and 5 meters, respectively). (D) The distribution of the increase in average path lengths (in minutes) for drivers when the process has been run up to $\tau$ = 5 shows that the most frequent trips (between 10 and 15 minutes) may see increases by a factor in the range between 1.5 and 3.}
\label{fig:montreal}
\end{figure}

\begin{figure}[h!]
\centering
	\includegraphics[width=.80\textwidth]{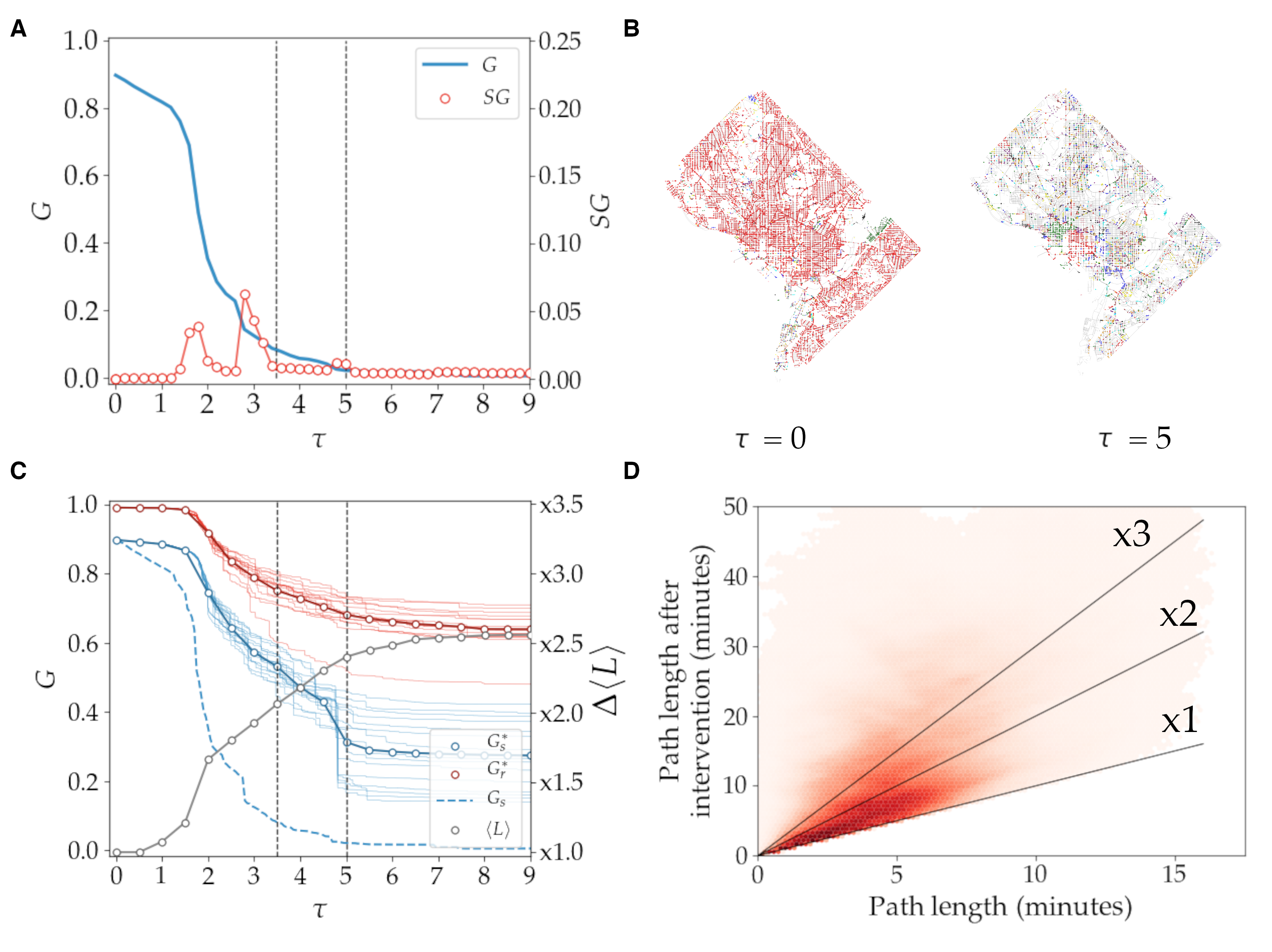}
\caption{{\bf Social distancing in Washington D.C.'s sidewalk network.}  (A) The evolution of G (solid blue), and the peak of SG (red circles) at $\tau$ = 2 meters indicate that the sidewalk network for this city presents a weak connectivity, and cannot withstand even a moderate requirement of social distance. (B) As for the case of New York City (Fig. 3 of the main text), Washington D.C. also presents a fragmented scenario at $\tau$ = 0, but the disconnected subnetworks are scattered in the outer periphery of the city. At $\tau$ = 5 only one component (in red) is appreciable in the center of the city. (C) The proposed ``Open Streets'' heuristic (solid blue) renders a large improvement with respect to the baseline, no-intervention scenario (dashed blue), with $G^*_s$ = 0.55 at $\tau$ = 3.5, and $G^*_s$ $\approx$ 0.35 at $\tau$ = 5. The cost of such improvement is on the road network, whose giant component $G^*_r$ is deteriorated by 20 to 25\% (at 3.5 and 5 meters, respectively). (D) The distribution of the increase in average path lengths (in minutes) for drivers when the process has been run up to $\tau$ = 5 shows that travel cost may go up by a factor of 3 or more at times, but the vast majority of trips (10 minutes or less) see a moderate increase by less than a factor of 2.}
\label{fig:dc}
\end{figure}

\begin{figure}[h!]
\centering
	\includegraphics[width=.80\textwidth]{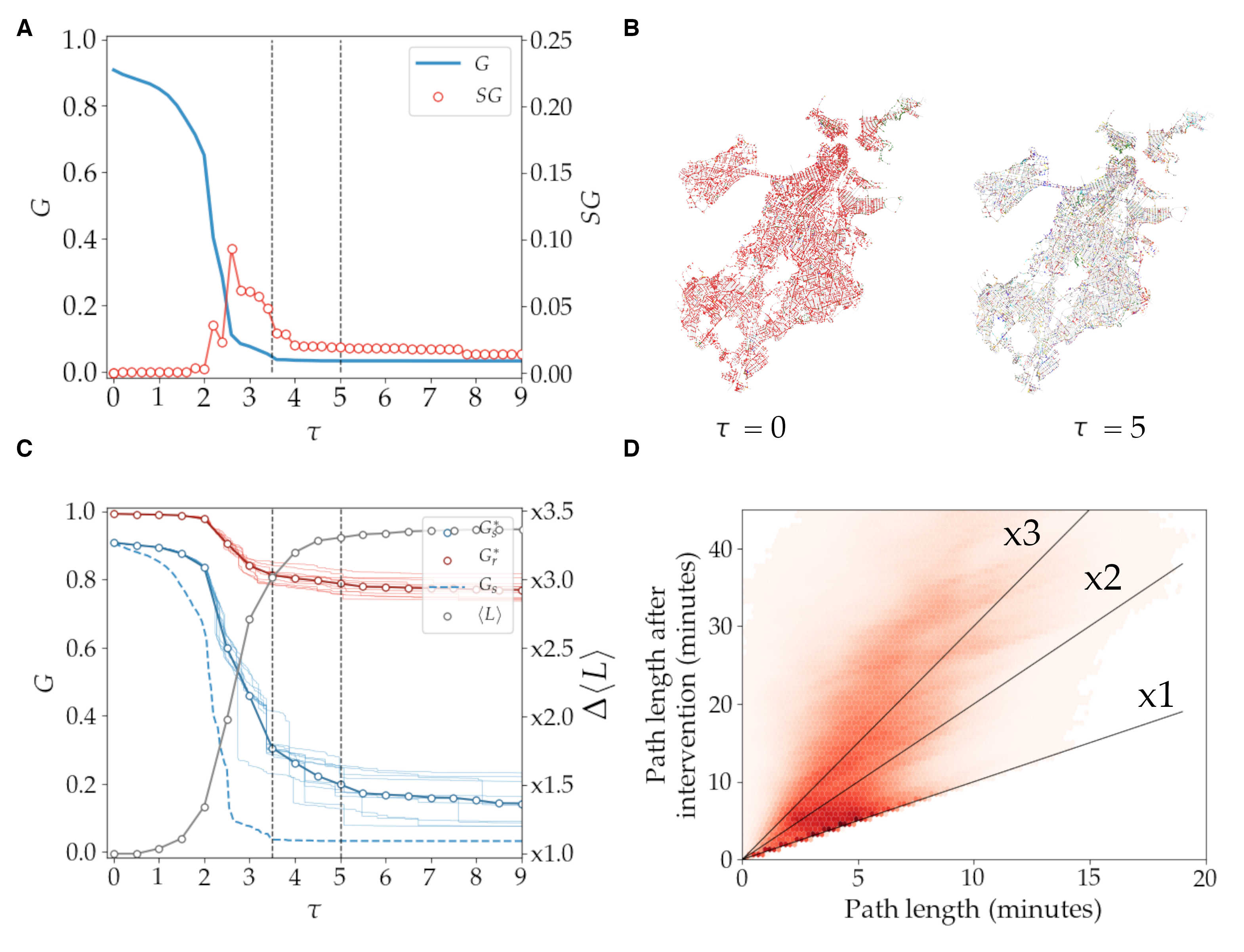}
\caption{{\bf Social distancing in Boston's sidewalk network.}  (A) The evolution of G (solid blue), and the peak of SG (red circles) at $\tau$ = 2.5 meters indicate that the sidewalk network for this city presents a moderately weak connectivity, and cannot withstand the requirement of social distance. (B) As for the case of New York City (Fig. 3 of the main text), Boston also presents a fragmented scenario at $\tau$ = 0, but the disconnected subnetworks are negligible in size (if compared to the giant component) and scattered in the outer periphery of the city. At $\tau$ = 5 there is no single component comparable to the size of the network. (C) The proposed ``Open Streets'' heuristic (solid blue) renders a notable improvement with respect to the baseline, no-intervention scenario (dashed blue), with $G^*_s$ = 0.35 at $\tau$ = 3.5, and $G^*_s$ $\approx$ 0.25 at $\tau$ = 5. The cost of such improvement is on the road network, whose giant component $G^*_r$ is deteriorated by 20. (D) The distribution of the increase in average path lengths (in minutes) for drivers when the process has been run up to $\tau$ = 5 shows that travel cost may go up by a factor of 3 or more at times, but the vast majority of trips (10 minutes or less) see a moderate increase by less than a factor of 2.}
\label{fig:boston}
\end{figure}

\begin{figure}[h!]
\centering
	\includegraphics[width=.80\textwidth]{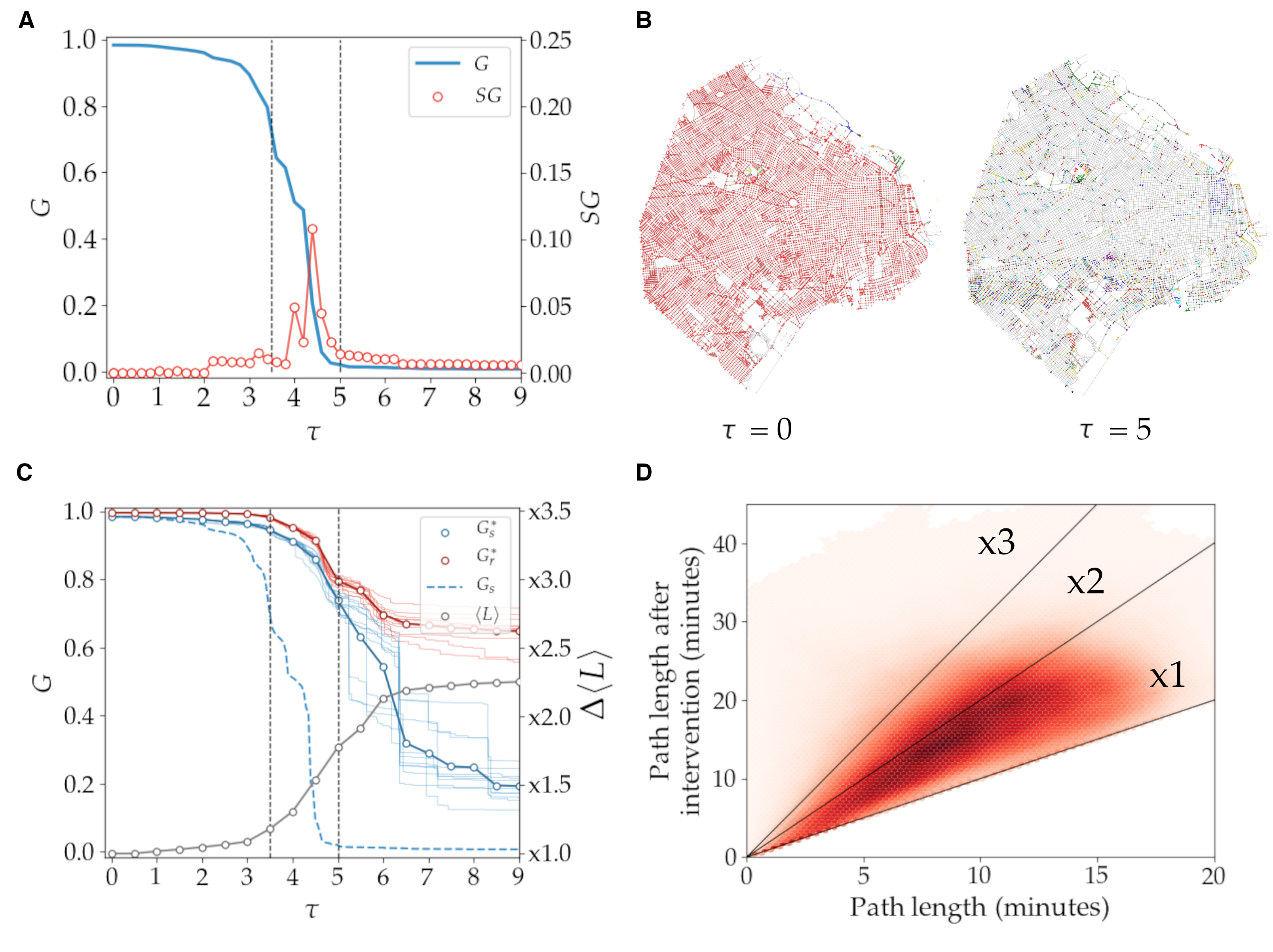}
\caption{{\bf Social distancing in Buenos Aires' sidewalk network.} (A) The evolution of G (solid blue), and the peak of SG (red circles) at $\tau$ = 4.5 meters indicate that the sidewalk network for this city presents a moderately strong connectivity, which may face the requirement of social distance at  $\tau$ = 3.5 (where G $\approx$ 0.7), but not at $\tau$ = 5, where the network has collapsed. (B) Unlike North American cities, Buenos Aires' sidewalk network is connected at $\tau$ = 0, but the disconnected subnetworks at $\tau$ = 5 are negligible in size (if compared to the size of the network). (C) The proposed ``Open Streets'' heuristic (solid blue) renders an excellent improvement with respect to the baseline, no-intervention scenario (dashed blue), with $G^*_s$ $\approx$ 0.95 at $\tau$ = 3.5, and $G^*_s$ $\approx$ 0.75 at $\tau$ = 5. The cost of such improvement is hardly noticeable on the road network, whose giant component $G^*_r$ is deteriorated between 5\% and 20\% (at 3.5 and 5 meters, respectively). (D) The distribution of the increase in average path lengths (in minutes) for drivers when the process has been run up to $\tau$ = 5 shows that travel cost may go up by a factor of 2 or more for some short trips (15 minutes or less), while longer ones see a moderate increase by less than a factor of 2.}
\label{fig:buenosaires}
\end{figure}

\begin{figure}[h!]
\centering
	\includegraphics[width=.80\textwidth]{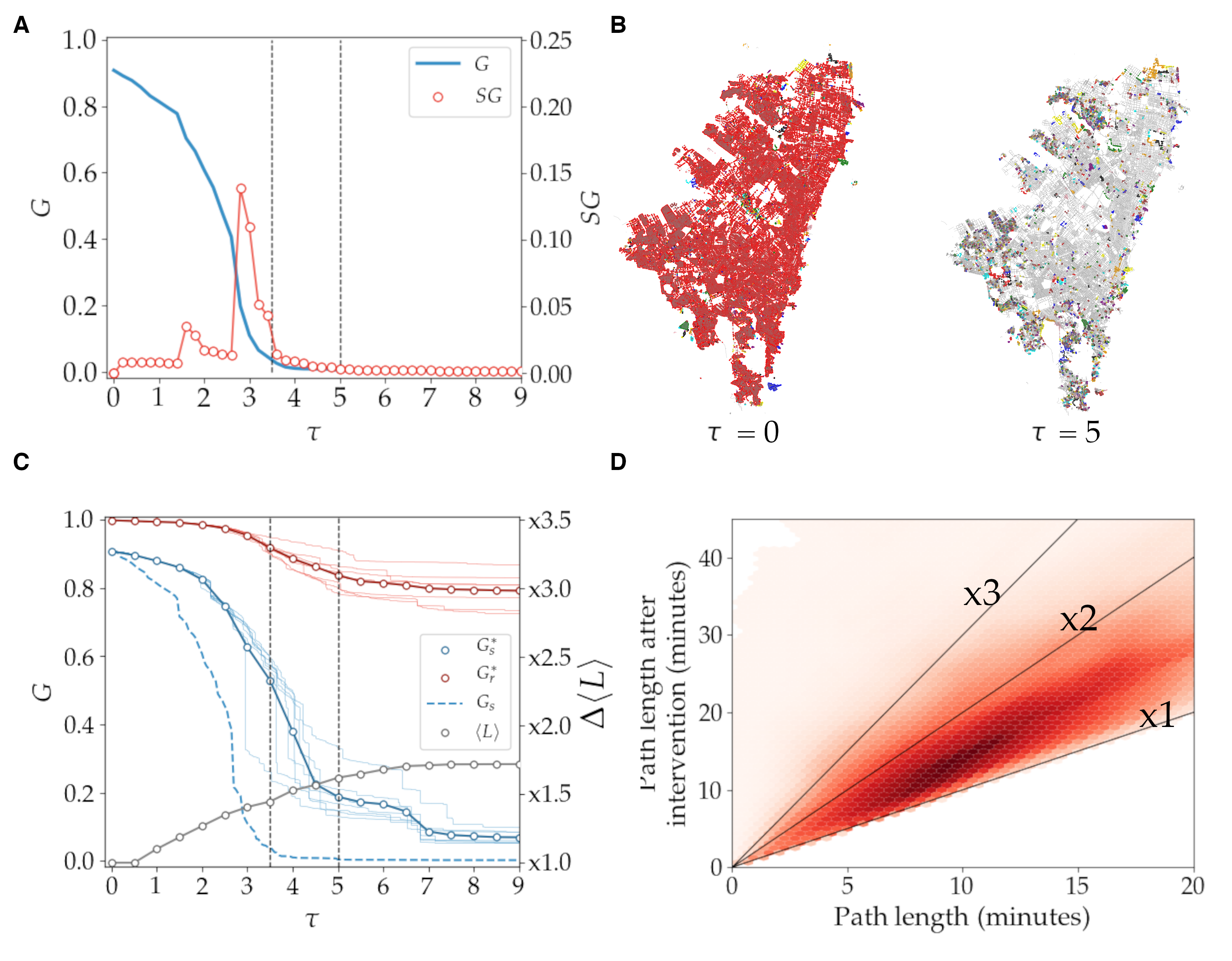}
\caption{{\bf Social distancing in Bogotá's sidewalk network.} (A) The evolution of G (solid blue), and the peak of SG (red circles) at $\tau$ = 2.8 meters indicate that the sidewalk network for this city presents a moderately weak connectivity, and cannot withstand the requirement of social distance. (B) Similar to North American cities, Bogotá's sidewalk network is already disconnected at $\tau$ = 0,  but the disconnected subnetworks are negligible in size (if compared to the giant component) and scattered in the periphery of the city. At $\tau$ = 5 there is no single component comparable to the size of the network. (C) The proposed ``Open Streets'' heuristic (solid blue) renders a good improvement with respect to the baseline, no-intervention scenario (dashed blue), with $G^*_s$ $\approx$ 0.55 at $\tau$ = 3.5, and $G^*_s$ $\approx$ 0.2 at $\tau$ = 5. The cost of such improvement is hardly noticeable on the road network, whose giant component $G^*_r$ is deteriorated between 10\% and 15\% (at 3.5 and 5 meters, respectively). (D) The distribution of the increase in average path lengths (in minutes) for drivers when the process has been run up to $\tau$ = 5 shows that travel cost may go up by a factor between 1 and 2 for all trips (15 minutes or less).}
\label{fig:bogota}
\end{figure}

\begin{figure}[h!]
\centering
	\includegraphics[width=.80\textwidth]{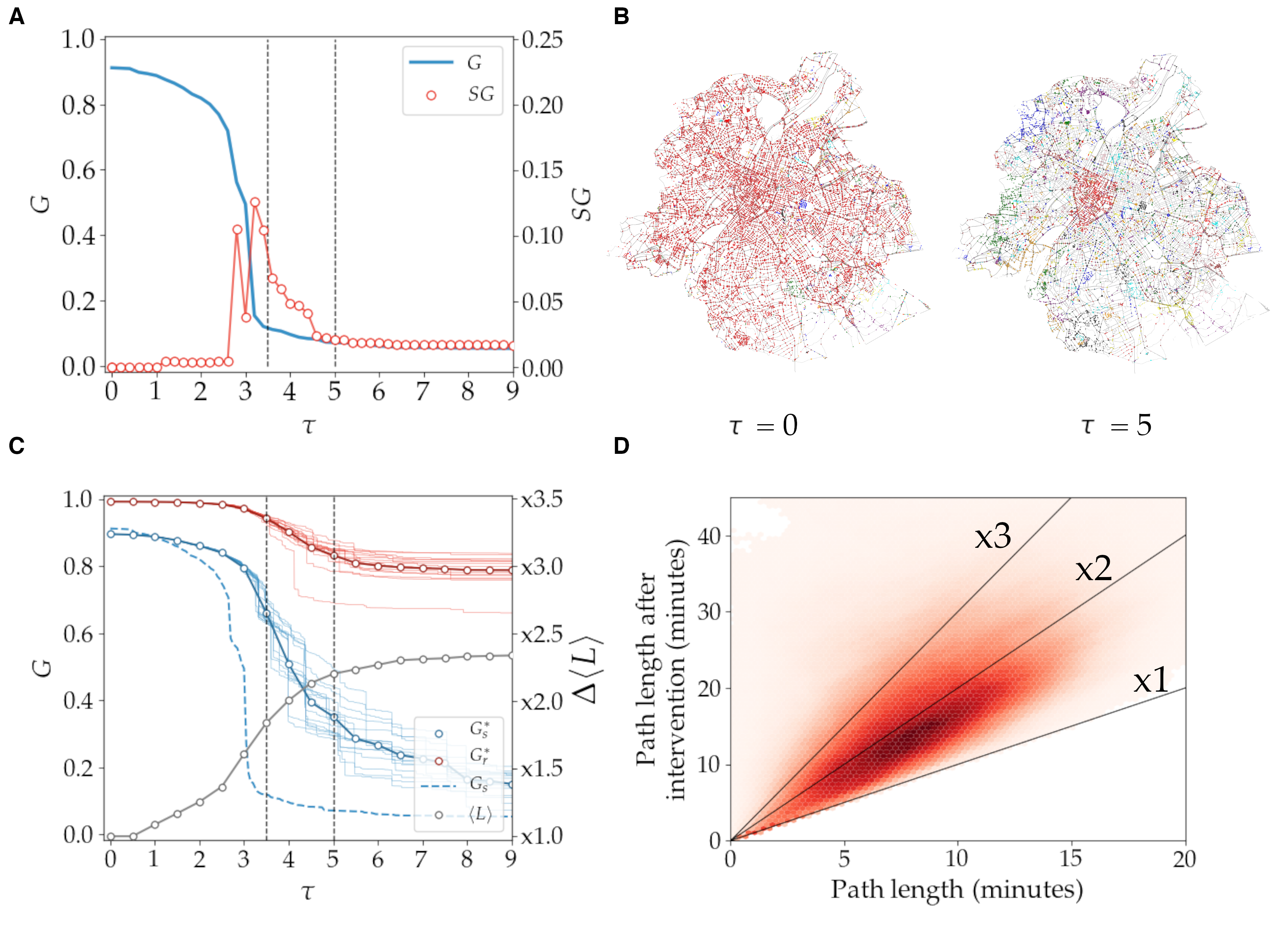}
\caption{{\bf  Social distancing in Brussels' sidewalk network.}  (A) The evolution of G (solid blue), and the peak of SG (red circles) around $\tau$ = 3 meters indicate that the sidewalk network for this city presents a moderately weak connectivity, and cannot withstand the requirement of social distance. (B) Similar to North American cities, Brussels' sidewalk network is already disconnected at $\tau$ = 0,  but the disconnected subnetworks are negligible in size (if compared to the giant component) and scattered through the city. At $\tau$ = 5 there is a single observable component comparable to the size of the network (red). (C) The proposed ``Open Streets'' heuristic (solid blue) renders a notable improvement with respect to the baseline, no-intervention scenario (dashed blue), with $G^*_s$ $\approx$ 0.65 at $\tau$ = 3.5, and $G^*_s$ $\approx$ 0.4 at $\tau$ = 5. The cost of such improvement is hardly noticeable on the road network, whose giant component $G^*_r$ is deteriorated between 5\% and 20\% (at 3.5 and 5 meters, respectively). (D) The distribution of the increase in average path lengths (in minutes) for drivers when the process has been run up to $\tau$ = 5 shows that travel cost may go up by a factor between 1 and 2 or more for the vast majority of trips (15 minutes or less).}
\label{fig:brussels}
\end{figure}

\begin{figure}[h!]
\centering
	\includegraphics[width=.80\textwidth]{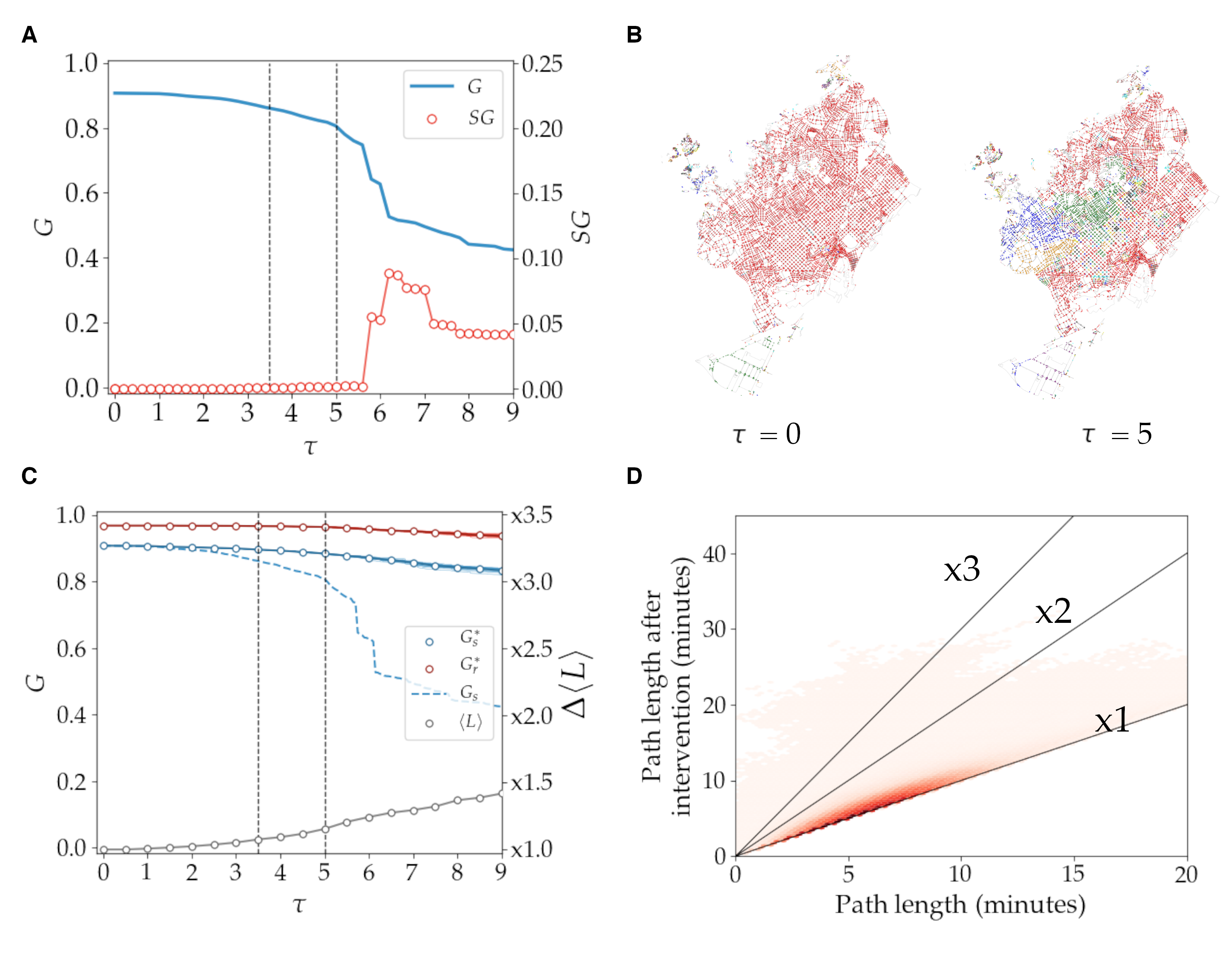}
\caption{{\bf Social distancing in Barcelona's sidewalk network.}  (A) The evolution of G (solid blue), and the peak of SG (red circles) at $\tau$ = 6.2 meters indicate that the sidewalk network for this city presents a very strong connectivity, which may face the requirement of social distance even up to $\tau$ = 5 (where G $\approx$ 0.8). (B) Similar to North American cities, Barcelona's sidewalk network is already disconnected at $\tau$ = 0, but the disconnected subnetworks are negligible in size (if compared to the giant component) and scattered at the periphery of the city. At $\tau$ = 5 there is still a clear giant component (red), and at least two other components visible to the naked eye (green, blue). (C) There is little room for improvement when the ``Open Streets'' heuristic (solid blue) is applied: $G^*_s$ presents almost the original value both at $\tau$ = 3.5 and $\tau$ = 5. The cost of such improvement is not noticeable on the road network, whose giant component $G^*_r$ is deteriorated by less than 5\% up to a width threshold of 5 meters. (D) Accordingly, the distribution of the increase in average path lengths (in minutes) for drivers when the process has been run up to $\tau$ = 5 shows that travel cost may almost remain constant for any travel time.}
\label{fig:barcelona}
\end{figure}

\begin{figure}[h!]
\centering
	\includegraphics[width=.80\textwidth]{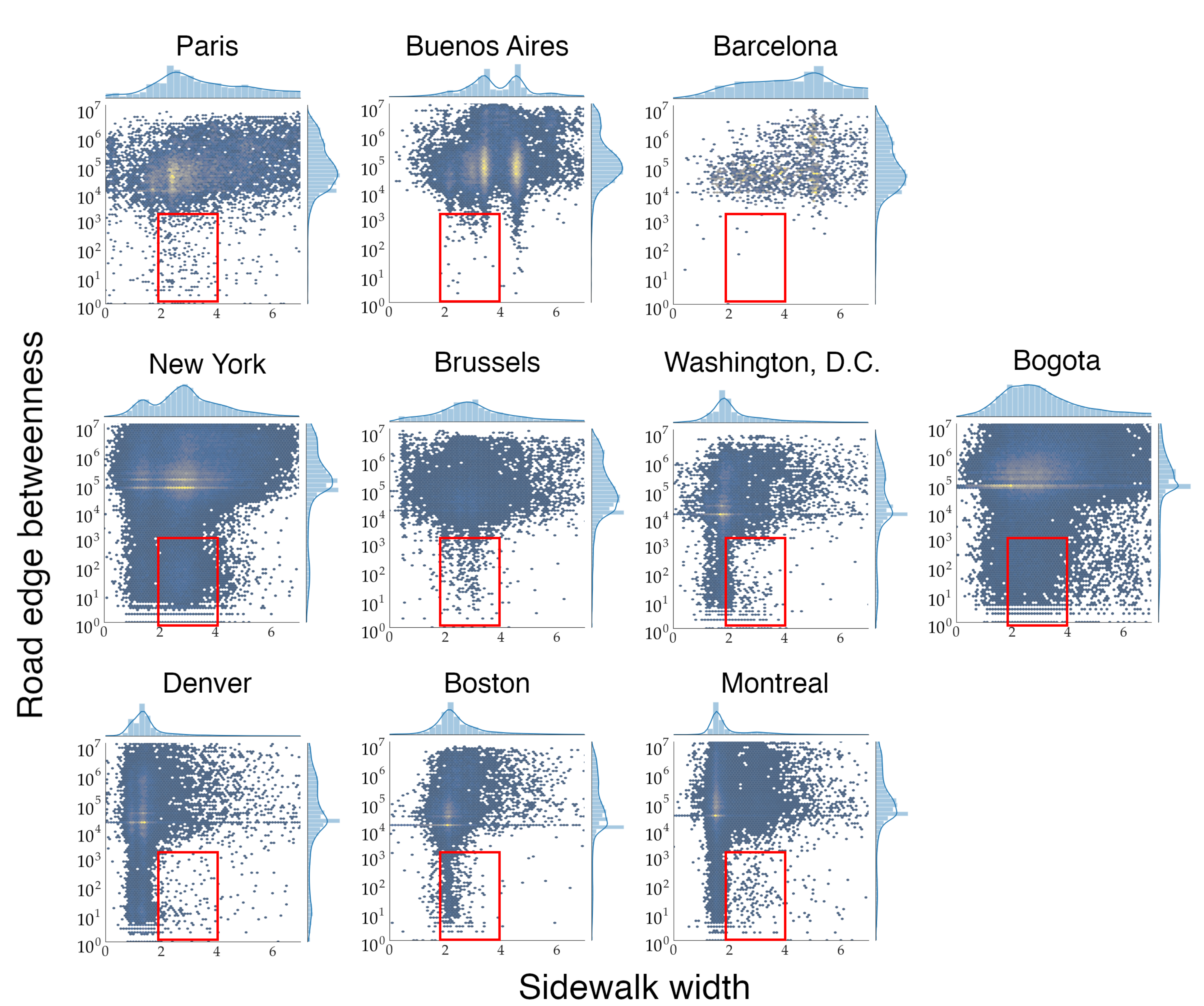}
\caption{{\bf Correlation of sidewalk width to road edge betweenness.} These scatter plots evince that there is no identifiable pattern that correlates sidewalk width with road centrality. Rather, each city presents its own and unique profile. If anything, it is possible to conclude that, typically, the widest sidewalks can be found in high-betweenness road segments. Such lack of correlation is expected, provided that there is no a priori necessity (nor obligation) for city governments to plan narrow/wide sidewalks as a function of road centrality. This variability explains the success (or lack thereof) for the proposed shared-effort heuristic, which demands to have a good amount of low betweenness roads (which are dispensable in this scheme) at middle ranges of sidewalk width. Note that we have informally sorted the scatter plots, from cities in which the heuristic was more (top)  to less (bottom) successful. The common trait for the least successful cities is precisely a lack of low-betweenness roadbeds in the range 2-4 meters (red square).}
\label{fig:widthbtw}
\end{figure}

%